# Super nucleation and orientation of poly (butylene terephthalate) crystals in nanocomposites containing highly reduced graphene oxide

Samuele Colonna[a], Ricardo A. Pérez[b], Haiming Chen[c], Guoming Liu[c], Dujin Wang[c], Alejandro J. Müller[b,d,*], Guido Saracco[a], Alberto Fina[a,*]

[a]Dipartimento di Scienza Applicata e Tecnologia, Politecnico di Torino, Alessandria 15121, Italy

[b]POLYMAT and Polymer Science and Technology Department, Faculty of Chemistry, University of the Basque Country UPV/EHU, Donostia-San Sebastián 20018, Spain

[c]Beijing National Laboratory for Molecular Sciences, CAS Key Laboratory of Engineering Plastics, Institute of Chemistry, Chinese Academy of Sciences, Beijing 100190, China

[d]IKERBASQUE, Basque Foundation for Science, Bilbao, Spain

*Corresponding author: alejandrojesus.muller@ehu.es , alberto.fina@polito.it

**Abstract**

The ring opening polymerization of cyclic butylene terephthalate into poly (butylene terephthalate) (pCBT) in the presence of reduced graphene oxide (RGO) is an effective method for the preparation of polymer nanocomposites. The inclusion of RGO nanoflakes dramatically affects the crystallization of pCBT, shifting crystallization peak temperature to higher temperatures and, overall, increasing the crystallization rate. This was due to a super nucleating effect caused by RGO, which is maximized by highly reduced graphene oxide. Furthermore, combined analyses by differential scanning calorimetry (DSC) experiments and wide angle X-ray diffraction (WAXS) showed the formation of a thick α-crystalline form pCBT lamellae with a melting point of ~250 °C, close to the equilibrium melting temperature of pCBT. WAXS also demonstrated the pair orientation of pCBT crystals with RGO nanoflakes, indicating a strong interfacial interaction between the aromatic rings of pCBT and RGO planes, especially with highly reduced graphene oxide. Such surface self-organization of the polymer onto the RGO nanoflakes may be exploited for the enhancement of interfacial properties in their polymer nanocomposites.

## 1. Introduction

Poly (butylene terephthalate) (PBT) is an engineering thermoplastic polymer used in a wide range of applications [1]. PBT can crystallize in two forms, namely the α-form and the β-form, both triclinic [2-4], but may also organize in a smectic liquid crystalline phase [5]. The α-form occurs when PBT is cooled from the melt, whereas the β-form is obtained upon uniaxial stretching (5 ÷ 15% strain) of PBT in the α-form. However, the β-form is not stable and after stress relaxation, the α-form is normally recovered [6-7]. The smectic phase is obtained by deformation of glassy PBT below room temperature, but is converted to the α-form upon heating [2, 5].

PBT melts at ~ 230 °C [8], and is characterized by a relatively high crystallization rate [9], good mechanical properties (except impact strength) [9-10], alkali resistance [11] and low melt viscosity [10]. Conventional PBT is synthetized by polycondensation of terephthalic acid with 1,4-butanediol [12-13]. However, cyclic butylene terephthalate oligomers (CBT) [14] may also be used as precursors for a catalyzed polymerization to produce linear poly (butylene terephthalate) (pCBT, to distinguish from conventional PBT) or cyclic poly (butylene terephthalate) (cPBT) [15]. The use of CBT as polymer precursors may be advantageous as compared to the traditional method, the former being an entropically driven athermal polymerization with no low-molecular-weight byproducts, occurring in mild conditions during extrusion processing, taking advantage of the low melting temperature (130- 170 °C) of the solid precursor, having an extremely low melt viscosity (~ 20 mPa s, at 190 °C) [15-17]. Furthermore, it is possible to modulate polymerization rate depending on catalyst type and concentration [16]. Typical number-average molecular weight values for pCBT are in the range 30000-50000 g mol$^{-1}$, with a polydispersity index in the range 2-3, depending on the catalyst type, polymerization time and temperature [15-16].

The crystallization of pCBT is strongly affected by polymerization temperature and CBT composition [18-20]. Indeed, Zhang et al. [19] polymerized CBT into pCBT for 30 minutes at selected polymerization temperatures in a DSC, observing that below 204 °C crystallization occurred during polymerization, leading to thick lamellar crystals with uniform crystal size distribution. When polymerization was performed at temperatures higher than 204 °C, crystallization and polymerization occurred separately, and above 212 °C only polymerization was observed. This behavior was reflected on the crystal size distribution, becoming wider above 204 °C polymerization temperature, with the appearance of double melting peaks, in the successive heating scan, related to the melting, re-crystallization and re-melting of thinner polymer crystals. Lehmann and Karger-Kocsis [18] carried out isothermal and non-isothermal crystallization experiments on pCBT, and observed different Avrami exponents, ($n \approx 2$ or 3) for pCBT obtained by distinct CBT mixtures. However, it is worth observing that in part of their isothermal experiments, the crystallization peak was partially overlapped with the transient signal of the DSC, an effect reported to affect by about 20% the estimation of Avrami parameters [21]. Wu and Jiang [20] studied pCBT crystallization by polarized optical microscopy and DSC and observed changes in the spherulitic shape of pCBT depending on the crystallization temperature with four different morphological features: (i) negative spherulites with a clear Maltese cross (usual spherulites) below 180°C, (ii) spherulites with a negative birefringence and mixed-type birefringence spherulites for crystallization temperature between 180 and 193 °C, (iii) mixed-type birefringence spherulites between 195 and 200 °C and (iv) highly disordered spherules for crystallization temperatures above 200°C. Finally, Zhang et al. [22] reported the co-existence of ring-banded and non-ring-banded morphology within one pCBT spherulite, with the non-ring-banded region showing axialite morphology.

The improvement of pCBT physical properties, *i.e.*, improvement of thermal stability, mechanical properties, electrical and thermal conductivity, etc., has been pursued by the polymerization of CBT in the presence of different types of nanoparticles, including carbon nanotubes (CNT) [23-24], organoclay [25-26], silica nanoparticles [27] and graphene-related materials (GRM) [28-34]. As far as we are aware, none of the reported works investigated in detail the crystallization of pCBT/GRM nanocomposites. However, a shift of the crystallization peak to higher temperatures, during non-isothermal DSC experiments, were reported for pCBT/reduced graphene oxide (RGO) [33] and pCBT/graphite nanoplatelets (GNP) [30, 32-33] nanocomposites. Furthermore, the addition of graphene-related materials to pCBT was reported to affect the melting behavior with a suppression of the double melting behavior of pCBT [12, 28, 32-33], thus suggesting the formation of more homogeneous crystal thickness distribution.

In the present work, we report the effect of both conventionally reduced graphene oxide and highly reduced graphene oxide on the crystallization of pCBT by means of differential scanning calorimetry (DSC), including advanced methods to study nucleation, self-nucleation and thermal fractionation of pCBT, in combination with wide-angle x-ray scattering (WAXS) to study crystalline structure and orientation.

## 2. Experimental

### 2.1. Materials

Cyclic butylene terephthalate oligomers [CBT100, Mw = $(220)_n$ g mol$^{-1}$, $n$ = 2-7, melting point = 130-170 °C] from IQ-Holding[*] (Germany) and butyltin chloride dihydroxide (96%, m$_p$ = 150 °C) were used as polymer precursor and ring-opening polymerization catalyst, respectively. A reduced graphene oxide (referred to as RGO) was used, having Surface Area ≈ 210 m$^2$/g, Oxygen content ≈ 3.2 at.%[†], Raman $I_D/I_G$ ≈ 0.88, $T_{Oxid}$ ≈ 558 °C[‡]. This product was synthetized by AVANZARE (Navarrete, La Rioja, Spain) according to a previously reported procedure [35]. The same RGO was annealed, in a closed graphite box, for 1 hour at 1700°C in a vacuum (p ≈ 50 Pa) oven (Pro.Ba., Italy) with graphite resistors. The product obtained after annealing is referred to as RGO_1700 and showed Raman $I_D/I_G$ ≈ 0.11, Oxygen content ≈ 0.4 at.%[†], $T_{Oxid}$ ≈ 750 °C[‡].

### 2.2. Nanocomposite preparation

In the present paper, pCBT nanocomposites containing 10 wt.% of RGO or RGO_1700 were prepared by a two-step procedure. In the first step, nanoflakes were premixed in acetone (99+% from Alfa Aesar, ~0.15 g mL$^{-1}$ CBT/acetone solution) obtaining a CBT/RGO mixture which, after solvent evaporation, was successively dried at 80 °C for 8h under vacuum. In the second step, the pulverized dried mixture was loaded into a co-rotating twin screw micro-extruder (DSM Xplore 15, Netherlands), and mixed for 5 minutes at 250 °C and 100 rpm. Then 0.5 wt.% of tin catalyst (calculated with respect to the oligomer amount) was added to the mixture and the process carried on for further 10 min to complete CBT polymerization into pCBT. The whole extrusion process was performed under inert atmosphere to avoid thermo-oxidative degradation and hydrolysis of the matrix.

pCBT + 50% RGO_1700 were prepared by solution mixing, 50 mg of RGO_1700 were added to ~ 150 mL of CHCl$_3$ (99.9+%, Sigma-Aldrich) and sonicated for 30 minutes in pulsed mode (30 seconds on and 30 seconds off, power set at 30% of the maximum) using an ultrasonication probe (Sonics Vibracell VCX-750). Then, ~ 8 mL of HFIP (99+%, Fluka) were added to the suspension and, later, 50 mg of pCBT were dissolved in the suspension for about 2 hours under vigorous stirring. When the dissolution of the polymer was completed, the solvent was evaporated, and the nanocomposite collected (in powder form), dried in vacuum at room temperature and, finally, stored in a glass vial.

---

[*] Distributor of products previously commercialized by Cyclics Europe GmbH
[†] XPS, O$_{1s}$ signal
[‡] Onset TGA plots in air, 10°C/min heating rate

## 2.3. Characterization

### 2.3.1. Transmission electron microscopy (TEM)

The nanocomposites morphology was observed by Transmission Electron Microscopy (TEM) with a TECNAI G2 20 TWIN (FEI) microscope, operating at an accelerating voltage of 200 kV in bright-field mode. Samples were sectioned with a Leica EMFC 6 ultramicrotome device at -25 °C equipped with a diamond knife. 300 mesh copper grids were used to support the ultrathin sections (~100 nm).

### 2.3.2. Differential scanning calorimetry (DSC)

Non-isothermal DSC scans, Self-Nucleation (SN) and Successive Self-Nucleation and Annealing (SSA) studies were performed in a DSC 8500 equipped with a Intracooler 3 cooling accessory (Perkin Elmer, USA). Isothermal crystallization experiments were carried out in a DSC Q20 equipped with a RCS 90 cooling system (TA Instruments, USA). Both instruments were calibrated with indium and zinc standards, and all the tests were performed with hermetically sealed aluminum pans under inert atmosphere ($N_2$) on dried samples (80 °C, ~ 100 Pa, overnight) to reduce hydrolysis of polymer.

*Non-isothermal DSC experiments*

Non-isothermal DSC experiments were carried out with 5.0 ± 0.5 mg samples in the range 25-270 °C using a heating rate of 20°C min$^{-1}$. Samples were heated up to 270 °C, held at this temperature for three minutes to erase thermal history, then a cooling scan was recorded down to 50 °C and finally a second heating run was performed till 270 °C. The crystallinity degree was calculated by assuming 140 J g$^{-1}$ as the heat of fusion of 100% crystalline PBT [36] and normalizing the enthalpy for the actual polymer content within the nanocomposites.

*Isothermal crystallization*

Isothermal crystallization tests were carried out with 2.5 ± 0.3 mg samples following the procedure recommended by Lorenzo et al. [21] Preliminary experiments were performed to ensure that no crystallization occurred during the rapid cooling to the selected $T_c$ range (see details in Ref. [21]). Samples were heated up to 260 °C for 1 minute to erase their thermal history. Then, samples were cooled at 40 °C min$^{-1}$ to the selected isothermal crystallization temperature, $T_c$, and held at this temperature for 40 minutes. Fitting to the Avrami equation was performed by the free Origin plug-in developed by Lorenzo et al. [21]

*Self-Nucleation studies*

The aim of self-nucleation (SN) is to produce self-nuclei by partially melting a "standard" crystalline state, taking into account that the ideal nucleating agent for any polymer should be its own crystal fragments or chain segments with residual crystal memory [37-39]. This technique was originally conceived for polymer solutions by Keller et al. [40], designed for DSC by Fillon et al. [37] and extensively exploited by Müller et al. [38]. Self-nucleation studies were carried out on 5.0 ± 0.5 mg samples, following this protocol:

(a) heating up to 260 °C (3 minutes isotherm at 260 °C) to erase thermal history and crystalline memory;
(b) cooling down to 25 °C at 20 °C min$^{-1}$ (1 minute isotherm at 25 °C) to create a standard crystalline state;
(c) heating up to a self-nucleation temperature, $T_s$, at 20 °C min$^{-1}$ and thermal conditioning at $T_s$ for 3 minutes;
(d) cooling scan from $T_s$ down to 25 °C at 20 °C min$^{-1}$ (followed by 1 minute isotherm at 25°C) to evaluate the effect of the thermal treatment on the crystallization behavior of pCBT;
(e) heating up to 260°C at 20 °C min$^{-1}$ to study the effect of the whole treatment on the melting of pCBT;
(f) repetition of step (b), (c), (d) and (e) at progressively lower $T_s$ values to identify the different *Domains* [37]

At the end of self-nucleation experiments, three possible *Domains* can be observed, as a function of the $T_s$: *Domain I* when $T_s$ is too high and complete melting of the sample occurs, *Domain II* when the melt retains some residual chain segmental orientation or crystalline memory (high temperature range) or some crystal fragments which cannot be annealed at the time spent at $T_s$ (low temperature range) and *Domain III* when $T_s$ is low enough to melt the material only partially and, simultaneously, anneal unmolten crystals during the conditioning for 3 minutes at $T_s$. Furthermore, defining the different *Domains* during SN experiments is crucial to obtain the starting $T_s$ for SSA tests.

*Thermal fractionation by SSA*

The aim of SSA technique is to perform an efficient thermal fractionation, *i.e.* to produce a distribution of lamellar crystals or thermal fractions by applying a series of temperature steps, for different times, to a crystalline material. This technique is performed with a conventional differential scanning calorimeter and was developed and reviewed by Müller et al. [38-39] Successive self-nucleation and annealing tests were performed on 2.5 ± 0.3 mg to compensate the heating rate increase. The following experimental protocol was adopted:

(a) heating up to 260 °C (3 minutes isotherm at 260 °C) to erase thermal history and crystalline memory;
(b) cooling down to 25 °C at 20 °C min$^{-1}$ (1 minute isotherm at 25 °C) to create a standard crystalline state;
(c) heating at 50 °C min$^{-1}$ up to the ideal self-nucleation temperature ($T_{s,ideal}$), defined as the minimum $T_s$ in Domain II, determined in SN experiments;
(d) holding at $T_{s,ideal}$ for 1 minute (this value represents the fractionation time, which was kept short to avoid possible degradation and was also constant for every fractionation step applied);
(e) cooling down to 25 °C at 50 °C min$^{-1}$ to crystallize the polymer after having been ideally self-nucleated;
(f) repetition of step (c), (d) and (e) at progressively lower $T_s$ values to produce annealing of unmolten crystals (*i.e.*, the thermal fractions) and self-nucleation of the molten polymer when the sample is cooled down. The fractionation windows, i.e., the difference in temperature between $T_{s,ideal}$ and $T_s$, was set at 5 °C and kept constant throughout the whole SSA experiment, determining the size of thermal fractions.
(g) Heating the sample up to 260 °C at 20 °C min$^{-1}$ to reveal the consequences of SSA fractionation.

### 2.3.3. Wide-Angle X-ray Scattering (WAXS)

WAXS measurements were performed on a Xeuss 2.0 SAXS/WAXS system (Xenocs SA, France). X-ray radiation (wavelength = 1.5418 Å) was produced by means of Cu-K$_\alpha$ radiation generator (GeniX3D Cu ULD) at 50 kV and 0.6 mA. Scattered signals were collected by a semiconductor detector (Pilatus 300 K, DECTRIS, Swiss) with a resolution of 487 x 619 pixels (pixel size 172 x 172 μm$^2$). Each room temperature WAXS pattern was obtained with 20 minutes exposure time. The one-dimensional intensity profiles were integrated from background corrected 2D WAXS patterns with an azimuthal angle range of 0-90°. Transmission geometry was adopted for *in-situ* measurements.

The temperature was controlled by a Linkam THMS600 hot stage (Linkam Scientific Instruments, UK). Heating and cooling rates for the measurement were set at 20 °C/min. Specimens were hold for 1 min at the selected temperature to stabilize the temperature, then WAXS were obtained with 5 min exposure times. The thermal protocol consisted of 4 heating steps (200 °C, 215 °C, 235 °C and 260 °C) and 9 cooling steps (250 °C, 240 °C, 230 °C, 220 °C, 210 °C, 200 °C, 190 °C, 180 °C and 150 °C). WAXS patterns were collected at room temperature (~ 30 °C) before the beginning and after the completion of the thermal protocol to evaluate structural changes which could occur while keeping the material at high temperatures for long times.

## 3. Results and discussion

### 3.1. Transmission electron microscopy (TEM)

Transmission electron microscopy was carried out to evaluate nanoparticle dispersion in pCBT nanocomposites. Representative TEM micrographs are reported in Figure 1. In both cases, homogeneous distribution of nanoflakes is observed, with polymer well infiltrated in the expanded structure of RGO and annealed RGO. These results are consistent with distribution and dispersion previously assessed in pCBT nanocomposites containing 5 wt.% of the same nanoparticles [28, 33].

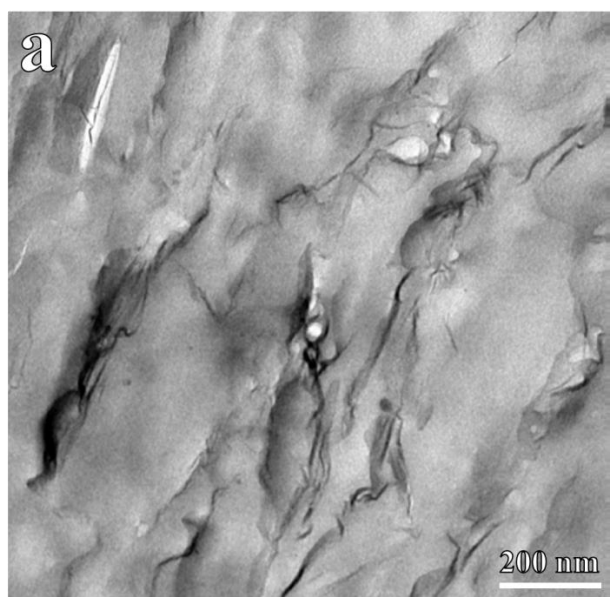 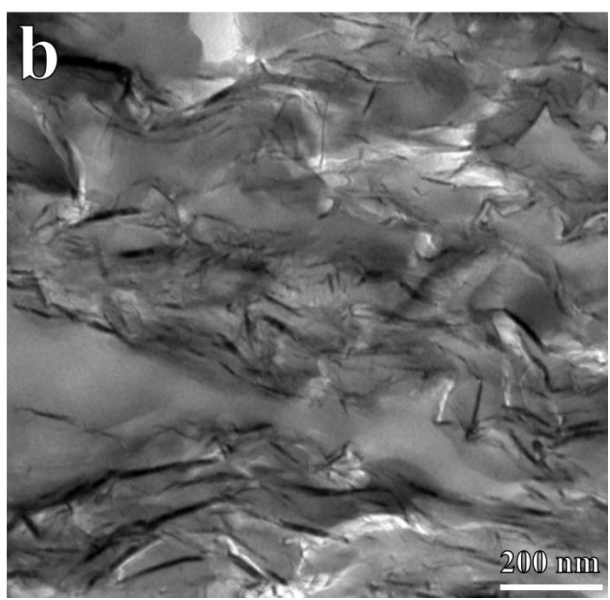

**Figure 1.** TEM micrographs for (a) pCBT + 10% RGO and (b) pCBT + 10% RGO_1700 nanocomposites.

### 3.2. Differential scanning calorimetry (DSC)

*Non-isothermal DSC experiments*

Non-isothermal DSC cooling scans, after erasing thermal history, and subsequent heating scans are reported in Figure 2, whereas the significant calorimetric parameters collected from these measurements are listed in Table 1.

After extrusion in the presence of the tin catalyst, none of the three materials exhibits traces of crystallization and melting typical for CBT oligomers, thus suggesting a high conversion of CBT into pCBT. Nevertheless, the absence of CBT crystallization/melting peaks is not sufficient to prove 100% conversion of CBT, conversion up to 97% was reported in literature for CBT polymerized under similar conditions (205°C, 3 min, same catalyst as in this work) [16].

In the presence of nanoflakes, the crystallization peak temperature shifts from ~ 190°C for pure pCBT up to ~ 201 °C and ~ 208 °C for pCBT + 10% RGO and pCBT + 10% RGO_1700, respectively, suggesting a strong nucleating effect of nanoflakes, which is typical for GRM in pCBT [28, 30, 32]. This effect on crystallization is reflected on the melting behavior of pCBT: neat pCBT exhibits two partially overlapping endothermic peaks, the first, at lower temperature (~ 217 °C), related to melting and crystallization of thin crystals, which subsequently re-crystallize and re-melt at higher temperatures (~ 223 °C), *i.e.,* in the second peak [41]. On the other hand, in nanocomposites only the higher temperature melting peak is observed; this is related to the formation of thicker crystals during cooling scans in presence of RGO nanoflakes, in agreement with Balogh et al.[32]

Comparing the effect of the different RGO, both crystallization and melting peaks for annealed RGO are located at higher temperatures, and appears to be narrower, thus suggesting a more efficient nucleation in presence of annealed RGO with the formation of thicker crystals. This more efficient nucleating action for pCBT + 10% RGO_1700, when compared with pCBT + 10% RGO, indicates that the surface structure of the nanoflakes (in terms of low defectiveness and high aromaticity of graphitic planes) plays a key role on the enhanced nucleation.

In pCBT + 10% RGO_1700, new crystallization and melting peaks appear, which are absent in pure pCBT, located at ~ 233 °C and ~ 250 °C during cooling and heating scans (see insets in Figure 2), respectively, with a calculated enthalpy of about 4 J g$^{-1}$. When carefully analyzing the DSC plots for pCBT + 10% RGO, similar peaks can also be detected. However, in the presence of untreated RGO, the peaks were located at slightly lower temperatures (~ 227 °C for crystallization and 247 °C for melting) and with a calculated enthalpy of about 1 J g$^{-1}$, further supporting the differences in pCBT crystallization in the presence of pristine vs. annealed RGO. Such high temperature crystallization and melting peaks have never been reported in pCBT literature during non-isothermal DSC scans, as far as the authors are aware. Only limited shifts of the melting peak of PBT to higher temperatures were reported in the literature [36, 42], after annealing PBT in DSC. Illers [36] annealed PBT for 850 h at 220 °C after quenching from the melt and observed a 10 °C shift, from 223 °C up to 233 °C, in the PBT melting peak. Yasuniwa *et al.* [42] submitted PBT to a stepwise annealing process, *i.e.,* consecutive isothermally annealing at progressively increasing temperatures, for an overall annealing time of 4140 minutes (~ 69 h) and observed the formation of two melting peaks, one located at ~ 223 °C and the other at 238.5 °C. In both papers, the authors conclude that the formation of the high temperature melting peak was related to an increase in the crystallite size of polymer lamellae. The high temperature melting peaks obtained in this work at 247 and 250 ºC are significantly higher than any value previously reported for annealed pCBT and could still be related to the formation of a thick stack of close to extended chain crystals almost extended within them. This explanation is consistent with reported values for pCBT equilibrium melting temperatures of 255.8 °C [15] and 257.8 °C [20].

The total crystallinity degree, calculated including both low and high temperature peaks for nanocomposites, is slightly affected by the presence of RGO, with an increase from 37% for neat pCBT up to 41% and 45% for pCBT + 10% RGO and pCBT + 10% RGO_1700, respectively, further confirming the influence of both types of RGO on the crystallization of pCBT.

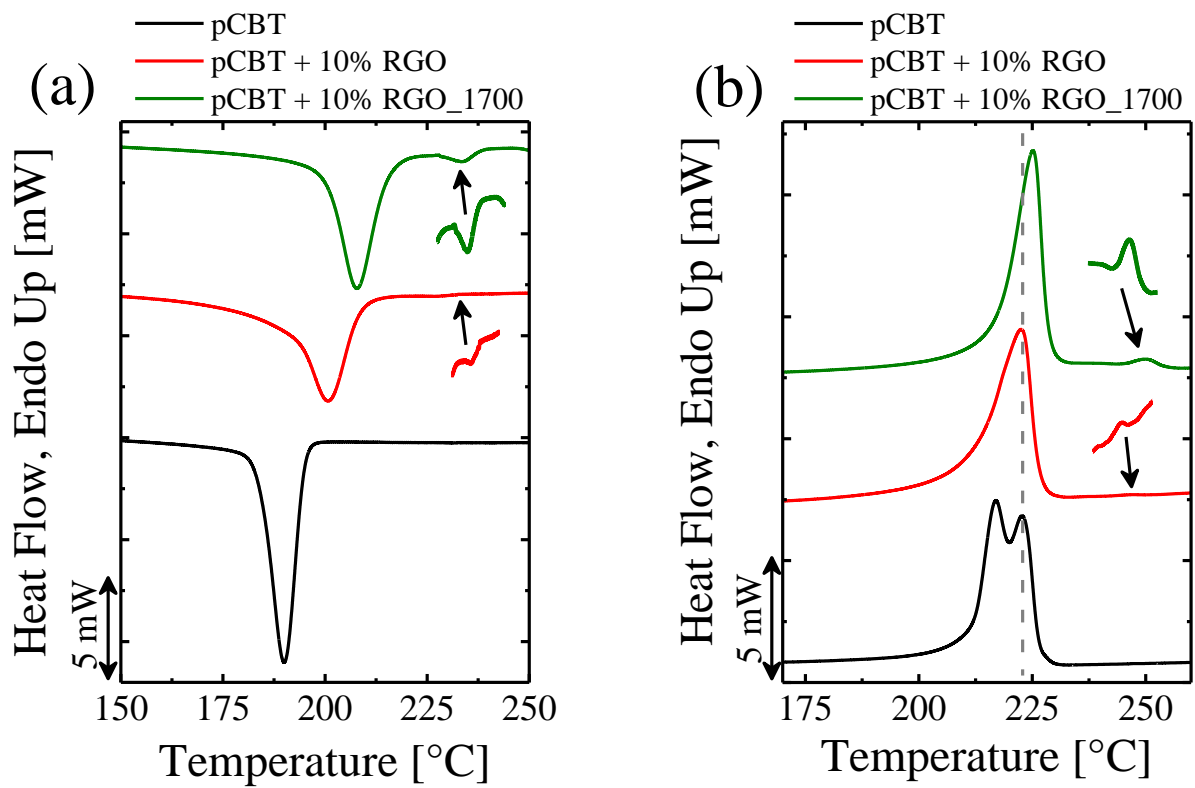

**Figure 2.** Standard DSC (a) cooling and (b) heating scans. The dashed gray line is reported for the sake of comparison

**Table 1.** Standard DSC results for pCBT and its nanocomposites

| Material | Cooling scans | | | | | Heating scans | | | | | |
|---|---|---|---|---|---|---|---|---|---|---|---|
| | $T_c$ | | $\Delta H_c$ | | $X_c$ | $T_m$ | | | $\Delta H_m$ | | $X_c$ |
| | [°C] | | [J g$^{-1}$] | | [%] | [°C] | | | [J g$^{-1}$] | | [%] |
| | $T_c^1$ | $T_c^2$ | $\Delta H_c^1$ | $\Delta H_c^2$ | | $T_m$ | $T_m^1$ | $T_m^2$ | $\Delta H_m^1$ | $\Delta H_m^2$ | |
| pCBT | 189.9 | - | 52 | - | 37 | 216.9 | 222.6 | - | 52 | - | 37 |
| pCBT+10%RGO | 200.7 | 227.0 | 56 | 1 | 41 | - | 222.5 | 246.6 | 56 | 1 | 41 |
| pCBT+10%RGO_1700 | 207.7 | 233.3 | 59 | 4 | 45 | - | 225.1 | 249.7 | 59 | 4 | 45 |

As nucleation effects are detected by the non-isothermal DSC results commented above, further studies were undertaken to elucidate the mechanisms of crystallization induced by the different RGO nanoflakes on pCBT, including isothermal crystallization and self-nucleation studies.

*Isothermal crystallization experiments*

Isothermal crystallization tests are used to measure the overall crystallization rate of a polymer (including both nucleation and growth). In this paper, the isothermal overall crystallization rate of pCBT and pCBT + 10 wt% RGO (including RGO and RGO_1700) was determined and the results are reported in Figure 3a as the inverse of the experimentally measured half-crystallization time (which is an experimental measure of the crystallization rate) *vs.* crystallization temperature.

Both pCBT nanocomposites need lower supercoolings to crystallize, in agreement with the results obtained by standard DSC experiments. However, the large difference in crystallization temperature range, between composites containing RGO_1700 and RGO, has to be highlighted. In fact, crystallization kinetics were found to be so different to make superposition of crystallization temperature ranges impossible. It is worth observing that for pCBT + 10% RGO_1700, only a limited number of data points were collected. In fact, at temperatures higher than 219°C, no crystallization peaks were observed, whereas below 218°C incomplete isothermal curves were recorded, indicating that crystallization started during cooling from the melt to the isothermal crystallization temperature. The increase of the crystallization rate for the nanocomposites is attributed to the nucleating effect of RGO, despite the possible role of changes in growth rate, which depend on the polymer molecular weight ($M_w$). Reductions in $M_w$ (with respect to neat pCBT) have been observed for nanocomposites prepared via ring-opening polymerization in the presence of nanoflakes for similar pCBT nanocomposites [34]. Indeed, it is well known that $M_w$ affects the crystallization rate of polymers, although the correlation is complex [43-44], and, so far no studies on the crystallization of pCBT with different $M_w$ have been reported in the literature.

Isothermal crystallization data were fitted to the Avrami theory (Figure 3b and Table SI1), which allows a simple and practical method to gain insight on pCBT crystallization. In fact, from the application of the Avrami theory, two main parameters are obtained: the Avrami index, *n*, and the overall crystallization rate constant, *k*, which contains contributions from both nucleation and growth [21]. The average Avrami index *n* (Figure SI3 and Table SI1) calculated for pure pCBT crystallization is about 2, which indicates the nucleation of instantaneous axialites [45]. In nanocomposites *n* values between 1.5 and 1.8 were obtained, suggesting that RGO does not alter the superstructural morphology of pCBT crystallization. Furthermore, results for the overall crystallization rate constant, raised to $n^{-1}$, as a function of crystallization temperature, displayed in Figure 3b, are consistent with experimental results obtained for the half-crystallization time, further proving the strong nucleating effect of RGO on pCBT.

The data in Figure 3a was fitted to the Lauritzen and Hoffman theory [46-47]. A good fit was obtained for both pristine pCBT and pCBT nanocomposites (Fitting parameters are listed in table SI2). It is noteworthy that, despite the limited amount of points available for pCBT + 10% RGO_1700, a reliable fit to the Lauritzen and Hoffman theory was obtained, thus indicating the consistency of the results. On the other hand, Lauritzen and Hoffman fitting (parameters are reported in table SI2) reveals that the presence of RGO leads to a reduction in the energy barrier required for nucleation and growth ($K_g$ values are proportional to this energy barrier), a decrease in the fold surface free energy ($\sigma_e$) and in the work required to fold chains ($q$). The largest effect is observed with annealed RGO.

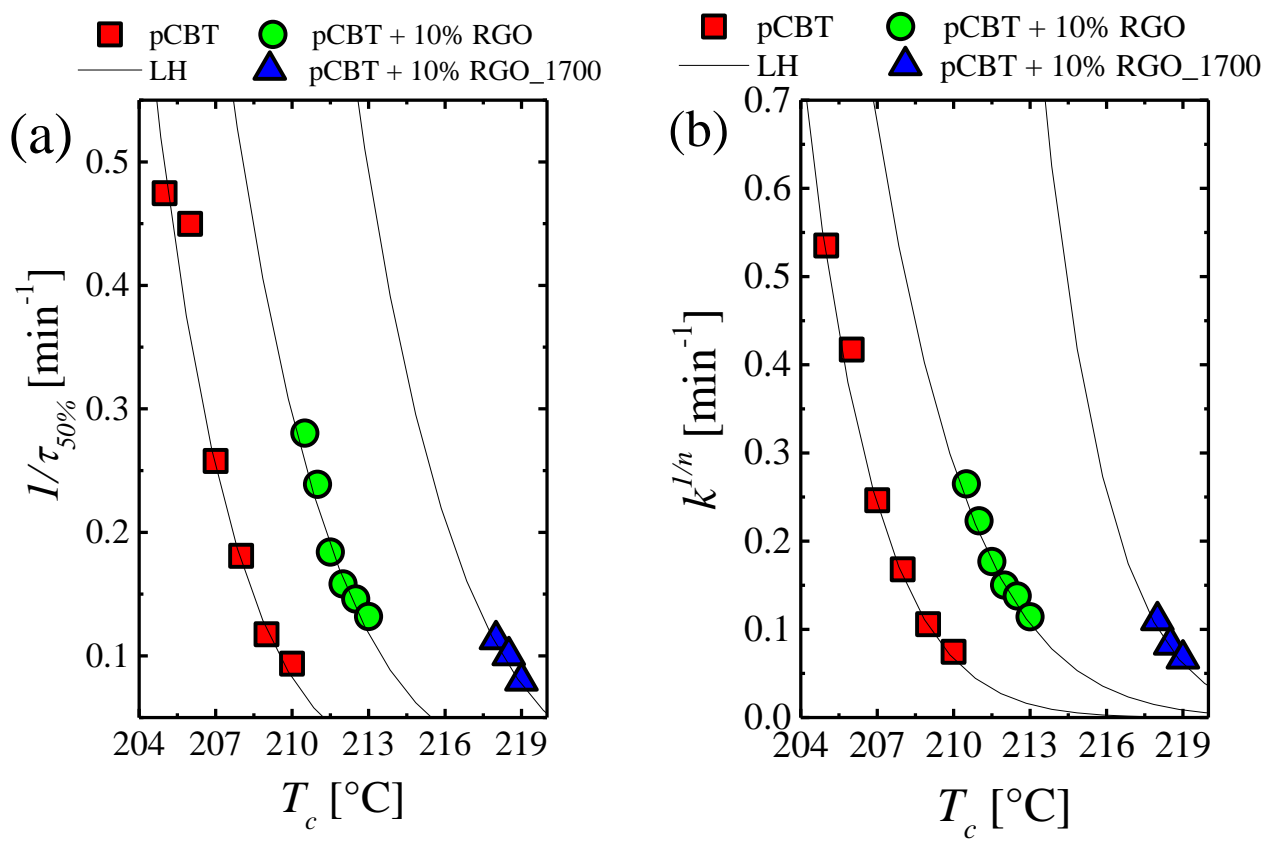

**Figure 3. (a) Overall crystallization rate ($1/\tau_{50\%}$) and (b) overall crystallization rate constant $k$ as a function of isothermal crystallization temperature for pCBT and pCBT/RGO nanocomposites.**

*Self-Nucleation and Nucleation Efficiency*

Following the results from standard and isothermal DSC experiments, which suggested the strong nucleation effect of RGO on the crystallization behavior of pCBT, a self-nucleation (SN) study was carried out to quantitatively assess the nucleation efficiency (NE) of RGO, as compared with pCBT self-nuclei.

In SN studies, the selection of the maximum temperature employed to erase thermal history requires careful optimization, based on the thermal stability of the polymer to be investigated. Therefore, a series of DSC cycles were performed to determine the stability of pCBT as a function of cycle number at the selected temperature, in the range 250-280°C, as reported and commented in the supporting information (Figure SI 2). As expected, the lower the temperature, the less severe the degradation was found. However, temperatures below 260°C cannot be used for the nanocomposites, owing to the presence of a small fraction of crystals with high stability, as commented above. Based on these constraints, the maximum temperature selected for thermal cycling in SN studies was 260°C.

Self-nucleation of neat pCBT was first studied to investigate the three *Domains* related to *I* the absence of self-nuclei, *II* the formation of self-nuclei and *III* the self-nucleation and annealing of unmolten pCBT crystals. Figure 4a displays DSC cooling plots following the heating ramp to a selected $T_s$ temperature, while Figure 4b reports the subsequent heating runs. For $T_s$ temperatures equal or higher than 231 °C, the $T_c$ temperature values were independent of $T_s$ (Figure 4a), indicating that the crystalline memory of pCBT was erased and crystals were completely molten. Furthermore, no clear alterations of melting profile (Figure 4b) were observed in the same temperature range. These indicates that neat pCBT is in *Domain I*, as defined by Fillon et al. [37].

In the $T_s$ temperature range 230-227 °C, the crystallization temperature gradually shifted to higher values (Figure 4a) upon decreasing $T_s$. Furthermore, changes in the melting behavior of pCBT (Figure 4b) were observed after treatment at $T_s$ 230-227 °C: the peak at lower temperatures, related to melting and recrystallization of imperfect crystals formed during cooling from $T_{max}$ [41], slightly shifted to higher temperatures, whereas the peak related to the main melting of pCBT remained unaltered. When $T_s$ = 227 °C, only one melting peak was observed, thus indicating that self-nuclei formed at that temperature allowed the production of thicker pCBT crystals during cooling from that $T_s$. The behavior observed in this $T_s$ range is characteristic of *Domain II*, where pCBT is nucleated by its own self-seeds, *i.e.* self-nucleation occurs. Indeed a $T_c$ shift to higher values is an indication of an increase in the nucleation density of pCBT. $T_s$ = 227 °C was therefore found as the ideal SN temperature, since it maximizes the nucleation density without altering the polymer melting behavior.

Finally, for $T_s$ equal or lower than 226°C a further shift of the crystallization peak to higher temperatures was observed (Figure 4a), whereas in the melting scans a small melting peak appeared at temperatures slightly higher than that of the main melting endotherm (indicated by an arrow in Figure 4b). The presence of this peak is related to the melting of annealed crystal fragments that did not melt at $T_s$ and annealed during the isothermal time spent at $T_s$, thus evidencing the behavior typical of *Domain III*. A schematic representation of $T_c$ vs. $T_s$, for neat pCBT, and the location of the different *Domains* is reported in Figure SI5a.

The efficiency of RGO as nucleating agents for pCBT was calculated by the following equation proposed by Fillon et al. [48]:

$$N.E. = \frac{T_{c,NA} - T_{c,pCBT}}{T_{c,max} - T_{c,pCBT}} \times 100 \qquad (1)$$

where $T_{c,NA}$ is the peak crystallization temperature of the polymer containing the nucleating agent (200.7 °C and 207.7 °C for pCBT + 10% RGO and pCBT + 10% RGO_1700, respectively), $T_{c,pCBT}$ is the peak crystallization temperature of neat pCBT after erasure of its crystalline memory (189.9 °C) and $T_{c,max}$ is the peak crystallization temperature (196.5 °C) obtained after pCBT was nucleated at 227 °C, identified as the ideal self-nucleation temperature.

Based on equation 1, the nucleation efficiency was calculated as *N.E.* = 164% and 270% for RGO and RGO_1700, respectively, thus indicating that RGO are significantly more efficient in nucleating pCBT with respect to its own self-nuclei. This effect has been termed super-nucleation [49] and, to the best of our knowledge, has never been reported in the literature for graphene-related materials. Actually, Dai et al. [50] reported a nucleating efficiency between 10 and 20 % for polypropylene nanocomposites containing 0.5 wt.% of GNP. Furthermore, the annealing of RGO nanoflakes demonstrated a dramatic effect on nucleation, leading to much higher nucleation efficiency when nanoflakes with low defectiveness are used.

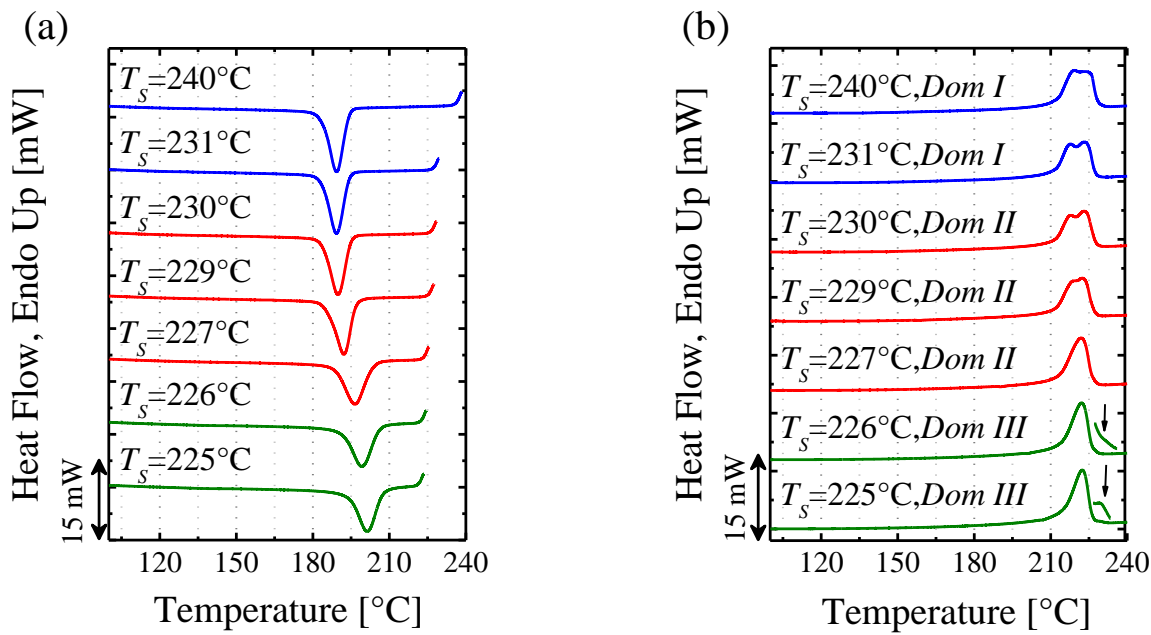

**Figure 4.** DSC (a) cooling scans from the indicated $T_s$ and (b) heating scans after cooling from the indicated $T_s$ for neat pCBT.

Beside the super-nucleation effect, it is also of interest to study how these nanoparticles affect the different *Domains* in the self-nucleation experiments. Results for pCBT + 10% RGO are reported in Figure 5a (DSC cooling plots for selected $T_s$ temperatures), Figure 5b (the subsequent heating runs) and Figure 5c (evolution of the different melting temperatures *vs*. $T_s$). Data for pCBT + 10% RGO_1700 are reported in Figure 6a (cooling plots for selected $T_s$ temperatures), Figure 6b (the subsequent heating runs) and Figure 6c (evolution of the different melting temperatures *vs*. $T_s$).

Comparing cooling and heating curves of the two nanocomposites, both RGO exhibited similar effects. Indeed, no significant shifts of the crystallization peak were observed, changing $T_s$ temperature, for both nanocomposites (refer to Figure SI5 for a comparison between $T_c$ *vs*. $T_s$ for both nanocomposites and that of pristine pCBT). Furthermore, also both the low and high temperature melting peaks did not exhibit any shift when varying $T_s$ but, in agreement with non-isothermal DSC experiments, the signal intensity for the high melting fraction was increased by the presence of thermally annealed RGO. However, it is worth observing the appearance of an additional broad endothermic peak, during heating scans, in the range 220-250 °C (see supporting information). The position of this peak ($T_{m,ann}$) appears to be directly affected by the selected self-nucleation temperature, as showed in Figure 5c and Figure 6c for pCBT + 10% RGO and pCBT + 10% RGO_1700, respectively. Indeed, the melting temperature of this peak increased with $T_s$, indicating annealing of pCBT matrix, which is typical of *Domain III*. This behavior in the presence of RGO could be expected, considering that self-nucleation experiments were carried out below the melting of the high temperature phase, which can play a key role in the nucleation and annealing of standard pCBT crystals.

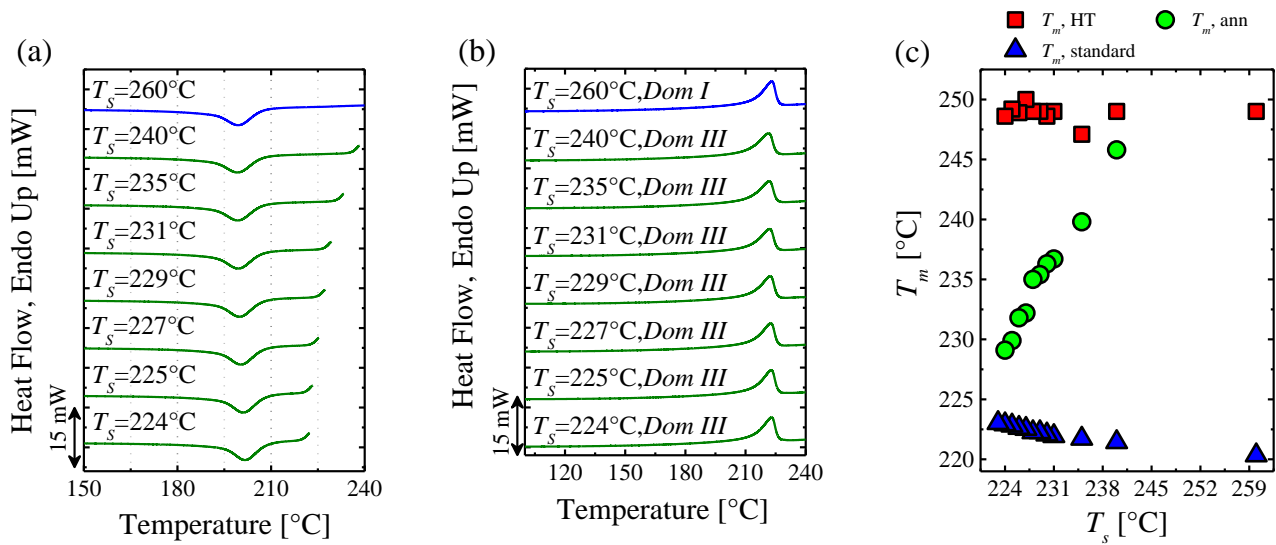

**Figure 5. DSC (a) cooling scans from the indicated $T_s$ and (b) heating scans after cooling from the indicated $T_s$ for pCBT + 10% RGO. (c) Evolution of the different melting temperatures *vs.* $T_s$. $T_m$,standard is the pCBT standard melting peak, $T_m$,HT is the melting peak related to the highly stable crystalline population, $T_m$,ann is the melting peak related to the annealed pCBT.**

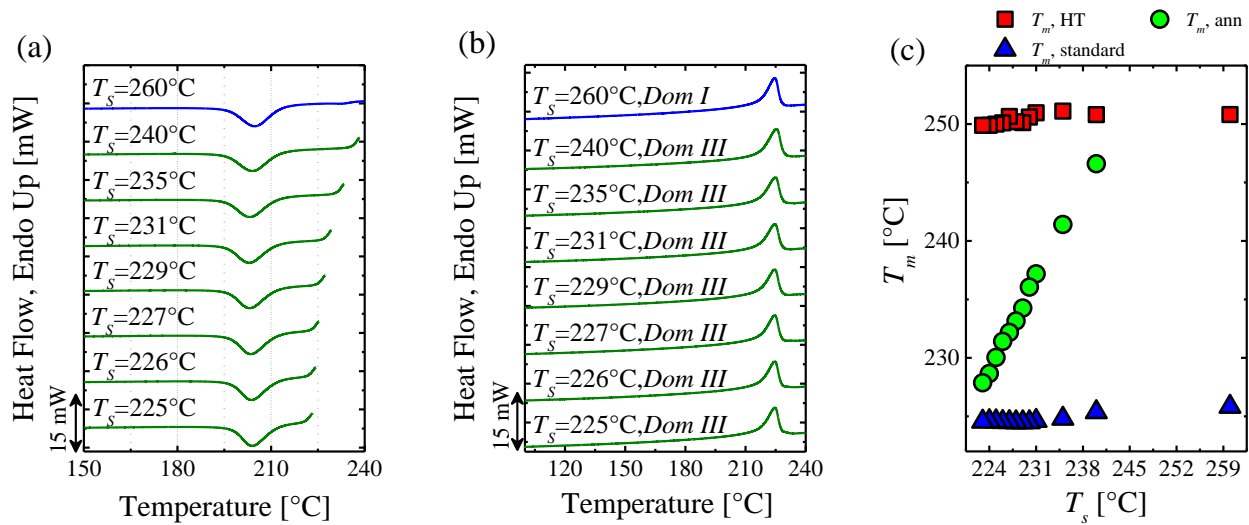

**Figure 6. DSC (a) cooling scans from the indicated $T_s$ and (b) heating scans after cooling from the indicated $T_s$ for pCBT + 10% RGO_1700. (c) Evolution of the different melting temperatures *vs.* $T_s$**

Self-nucleation experiments on neat pCBT, reported above and discussed, showed the presence of the three *Domains* defined by Fillon *et al.* [37] with the ideal self-nucleation temperature $T_s$ = 227 °C. On the other hand, the presence of RGO drastically changed the pCBT behavior in SN tests, with annealing occurring even when the standard pCBT crystals should be molten, thus indicating that the polymer, in the selected temperature range, is in *Domain III*. This behavior could be related to the presence of the highly stable crystalline population that melts at temperatures above 240 ºC.

*Successive Self-Nucleation and Annealing*

Successive self-nucleation and annealing (SSA) is a thermal fractionation technique designed to produce a distribution of lamellar crystals or thermal fractions. This technique is particular sensitive to the presence of defects in the chains, therefore it is particularly valuable for the study of copolymers, branched polymers, stereo-defects, etc [39]. On the other hand, SSA can be exploited also for the characterization of linear polymers, where fractionation occurs only for chain length differences, even if the fractionation is less efficient [51]. Thermal fractionation experiments on pCBT were performed setting as first $T_s$ temperature the $T_{s,ideal}$ determined in self-nucleation experiments, *i.e.*, $T_s$ = 227 °C. The thermal protocol consisted in seven $T_s$, from 227°C down to 197°C. Despite Müller *et al*. [38-39] suggested 5 minutes as ideal fractionation time at $T_s$, in the present paper 1 min was employed to limit the thermal degradation of the polymer matrix during SSA experiment.

For neat pCBT, DSC heating scan after completion of SSA and the second heating measured by non-isothermal DSC experiments are reported in Figure 7. After SSA, a series of melting peaks are usually produced, depending on the effectiveness of the thermal treatment to separate fractions. In this case, as pCBT is a linear polymer the fractionation produced is not well resolved (i.e., the melting peaks are not well separated from one another). The shape of DSC curve drastically changed after thermal fractionations and the distribution of melting points produced by SSA only reflects melting fractions with no recrystallization during the scan (as recrystallization processes or reorganization during the scan are intrinsically avoided by annealing effects induced by SSA). The thermal cycles applied during SSA produce much thicker lamellae as effective annealing of the material is produced, hence a higher melting point.

The protocol for SSA thermal fractionation of nanocomposites was slightly changed with respect to that of neat pCBT, owed to the presence of the high melting phase. Indeed, twelve $T_s$ temperatures (indicated by vertical lines) were selected, starting from 252°C down to 197°C, still assuming $T_s = 227$ °C as $T_{s,ideal}$ (segmented blue vertical lines). Results for pCBT + 10% RGO are reported in Figure SI6a and Figure SI6b, whereas corresponding plots for pCBT + 10% RGO_1700 are reported in Figure SI6c and Figure SI6d. For both nanocomposites, thermal fractionation of the polymer matrix was observed for the main melting peak of pCBT, as well as for the high-temperature melting peak. This is a further proof that the high temperature crystalline population is related to real polymer crystals, which can be annealed and fractionated. Finally, it is worth observing that after thermal fractionation, in pCBT + 10% RGO_1700 the highest melting peak temperature is centered at ~ 253 °C, which is once again close to the equilibrium melting temperature estimated in the literature for neat pCBT [15, 20]. This supports the hypothesis that the highest melting point fraction correspond to the melting of extended chain crystals, especially in the presence of RGO with low defectiveness and oxidation, which appears to have higher interaction with polymer chains.

The thickness of pCBT lamellae was roughly estimated by the Gibbs-Thomson equation:

$$T_m = T_m^0 \left(1 - \frac{2 \cdot \sigma_e}{l \cdot \rho_c \cdot \Delta H_f^0}\right)$$

where $T_m$ is the melting temperature, $T_m^0$ represent the equilibrium melting temperature, $\sigma_e$ is the surface fold free energy, $l$ is the lamellar thickness, $\rho_c$ is the density of the crystalline phase (1.397 g cm$^{-3}$ [52]) and $\Delta H_f^0$ is the enthalpy of fusion of a completely crystalline sample (140 J g$^{-1}$ [36]). For the calculation of the lamellar thickness, the end melting point of the high temperature melting fraction was selected as the equilibrium melting temperature, with a value $T_m^0 \approx 257.6$ °C, which is close to the 257.8 °C estimated by Wu et al. [20], whereas a surface fold free energy value $\sigma_e = 57$ erg cm$^{-2}$ [53] was used. It is worth observing that for PBT, $\sigma_e$ values reported in the literature range from 34 to 85 erg cm$^{-2}$ [10, 53-54].

The Gibbs-Thomson equation was employed to convert temperature into lamellar thickness values, thus plotting SSA plots *vs.* lamellar thickness (Figure 8). Results show that thermal fractionation leads to unimodal lamellar thickness distribution ranging within 3 and 6 nm with the maximum centered at about 4.8 nm for pCBT and pCBT + 10% RGO_1700 and 4.6 nm for pCBT + 10% RGO. These results are in agreement with data reported in literature for poly(butylene terephthalate) [5, 8], where the lamellar thickness was shown to be dependent on the isothermal crystallization temperature. Konishi et al. [5]

performed SAXS measurements on PBT isothermally crystallized at ~ 188 °C for 35 min and estimated a lamellar thickness of 5.2 ± 1.2 nm, while Hsiao *et al.* [8] estimated a crystalline lamellar thickness of ~ 6.0 and ~ 8.0 nm for PBT ($M_W$ = 45000 g/mol) isothermally crystallized at 130 and 175 °C, respectively. Furthermore, in the same work higher lamellar thicknesses were obtained for PBT with higher molecular weight, indicating a correlation between $M_W$ and *l*. Zhang *et al.* [22] measured lamellar thickness of pCBT lamellae in spherulites through a digital image processing software and measured about 12 nm thickness. However, it is worth noting that a layer of platinum was sprayed on the top of spherulites, thus possibly leading the authors to an overestimation of lamellae thicknesses.

Focusing on the high temperature melting fraction (Figure 8b), no endotherm peaks are visible for pCBT and pCBT + 10% RGO, as expected. The DSC final heating run after SSA for pCBT + 10% RGO_1700 exhibits two peaks, centered at ~ 20 and ~ 32 nm lamellar thickness (Figure 8b, see vertical segmented lines), respectively, thus indicating the formation of lamellae with a thickness approximately 4 and 6 times higher than the "standard" pCBT lamellae. The length of one repeating unit, supposing an all trans conformation for pCBT chains, can be roughly estimated as 1.417 nm, with a molecular weight $M_w$ = 220 g mol$^{-1}$. Assuming completely extended chains, in thicker lamellae, we can estimate they were formed by about 23 repeating units, thus $M_w \approx$ 5000 g mol$^{-1}$, for chain length L = 32 nm. Given the average viscosity molecular weight for similar nanocomposites was previously reported in the range 20000-30000 g mol$^{-1}$ [34], it can be speculated that lower molecular weight fraction may organize into Extended Chain Crystals (ECC), especially in the presence of low defective RGO.

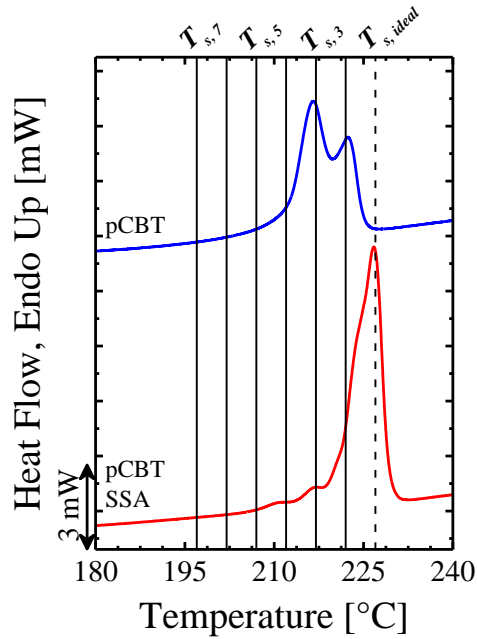

**Figure 7.** DSC heating scans for pCBT before (blue curve) and after (red curve) SSA thermal fractionation. The solid vertical lines represented the values of $T_s$ temperature employed for thermal fractionation while the dashed vertical line indicates the $T_{s,ideal}$ for pCBT.

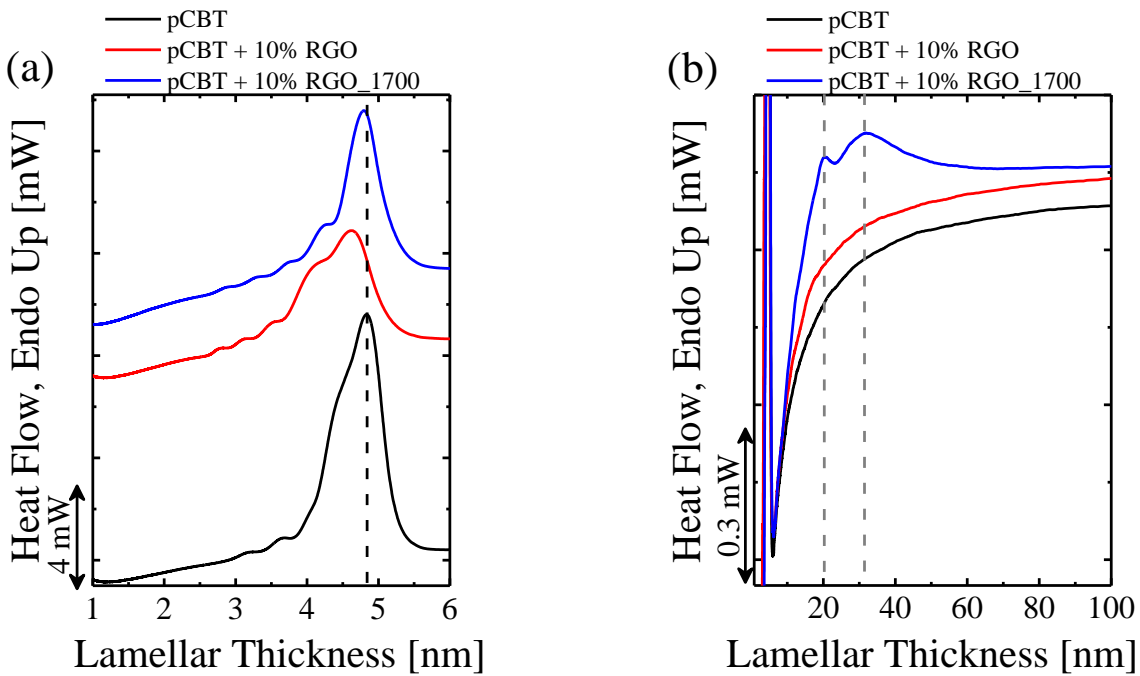

**Figure 8.** Heat flow *vs.* lamellar thickness for pCBT and its nanocomposites after SSA thermal fractionation. Effect of thermal fraction on (a) the standard and (b) high temperature melting fractions.

### 3.3. Wide Angle X-ray Scattering (WAXS)

While DSC experiments, revealed the presence of a small high melting temperature crystal population, which can be annealed and fractionated, such measurements cannot provide any information on the crystalline structure of this new high temperature crystal fraction. For this reason, WAXS experiments were performed first at room temperature and then heating specimens above the main pCBT melting temperature, aiming at the detection of the diffraction pattern from the highly stable crystalline fraction.

WAXS patterns collected via transmission geometry on pCBT, pCBT + 10% RGO and pCBT + 10% RGO_1700 are presented in Figure 9. Independently on the presence of RGO nanoflakes, all the WAXS patterns revealed peaks centered at diffraction angles, $2\Theta$, 8.9° (001), 16.0° (0$\bar{1}$1), 17.2° (010), 20.5° ($\bar{1}$11), 23.2° (100), 25.3° (1$\bar{1}$1), 29.2° (101) and 31.2° (1$\bar{1}$2), thus indicating that pCBT crystallized in its alpha crystalline form [3, 55-56]. Reflections indexing was estimated based on the atomic positions for pCBT reported by Yokouchi et al. [4]. However, some different indexing can be found in the literature [55-56], owing to partial overlapping of different reflections which can lead to an ambiguous interpretation of data. The appearance of a shoulder at $2\Theta \approx 26.2°$ in pCBT nanocomposites, was related to the (002) reflections of graphite [35], which is expected in the presence of RGO nanoflakes with thickness in the range of several nanometers. WAXS measurements performed with the incident X-rays perpendicular to the compression direction (Figure 9a) show a tiny signal related to the presence of RGO, in both nanocomposites, and an almost isotropic 2D patterns (pCBT + 10% RGO_1700 reported as example in Figure 10a).

WAXS patterns collected setting the incident X-rays parallel to the compression direction (Figure 9b) displayed a more intense peak at ~ 26.2°, thus evidencing a preferential orientation of nanoflakes parallel to the specimen surface, which is expected given their high aspect ratio. Furthermore, a clear anisotropy is observed for pCBT signals in the nanocomposites, with polymer chains preferentially aligned parallel to the RGO sheets, especially in the case of pCBT + RGO_1700. Orientation is observable by differences between the (100) and (1$\bar{1}$1) reflections in patterns collected perpendicular (Figure 9a) and parallel (Figure 9b) to the compression direction and on the 2D pattern collected parallel to the compression direction (Figure 10b). Analyzing the intensity distribution of the main reflections *vs.* the azimuthal angle for pCBT + 10% RGO_1700 (supporting information, Figure SI7) it appears clear that (100), (1$\bar{1}$1) and ($\bar{1}$11) reflections orient parallel to the (002) reflection of graphitic materials. In particular, (100) and ($\bar{1}$11) planes show similar plane orientations even if less pronounced with respect to that of (1$\bar{1}$1) planes, which are the most oriented pCBT planes. On the other hand, (010), (001) and (0$\bar{1}$1) reflections orient perpendicularly with respect to RGO planes, even if they exhibit an overall lower orientation. It is interesting to observe that (1$\bar{1}$1) planes are almost parallel to the benzene rings of pCBT chains, thus suggesting orientation related to a π-π interaction between pCBT and RGO surface. This interaction is maximized in presence of annealed RGO, thus confirming a higher degree of self ordering of pCBT macromolecules onto the lower defectiveness of RGO_1700, compared to pristine RGO [35].

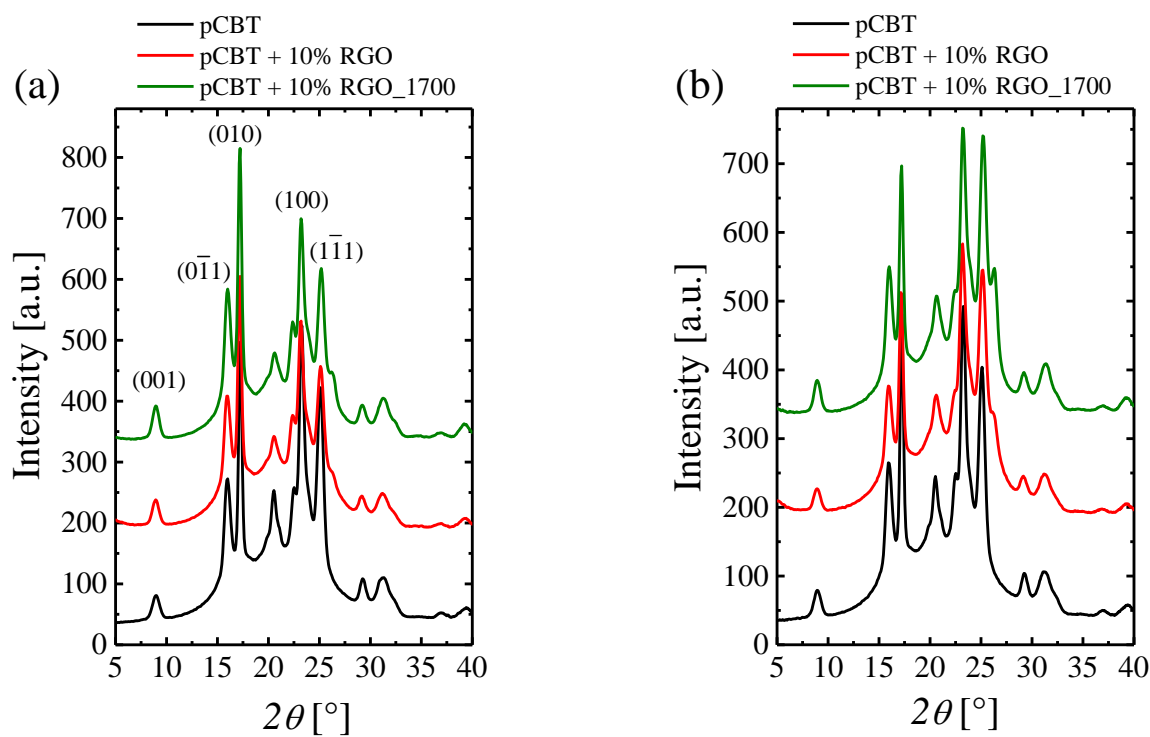

**Figure 9. WAXS patterns measured via transmission geometry on pCBT, pCBT + RGO and pCBT + RGO_1700. WAXS measured (a) perpendicular and (b) parallel to the compression direction.**

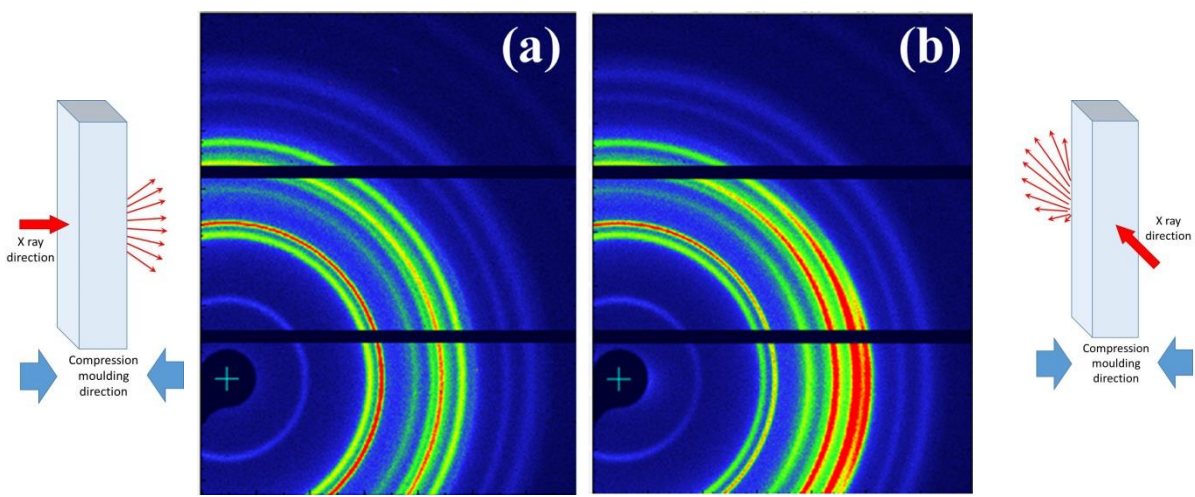

**Figure 10.** 2D WAXS patterns measured via transmission geometry on pCBT + 10% RGO_1700 (a) parallel and (b) perpendicular to the compression direction. A schematic of pattern collection is reported on below each 2D WAXS patterns. 2D diffraction patterns of pCBT and pCBT + 10% RGO are reported in Figure SI 8 and Figure SI 9.

Besides marked anisotropy in the polymer nanocomposites, neither peaks shifts nor new peaks were found in the diffraction pattern of pCBT nanocomposites vs. neat pCBT, thus suggesting the high temperature melting/crystallization fraction is not related to a new crystalline phase. To gain more insight on the crystalline organization of the high melting point fraction, *in situ* variable temperature WAXS measurements were carried out, aiming at the melting of the pCBT main crystal fraction while preserving the highly stable crystals. Variable temperature *in situ* WAXS patterns collected for pure pCBT are reported in Figure 11. Starting from the top of the figure, the four red curves represent the diffraction patterns collected at the reported temperature during heating. While the diffractogram is fully consistent with the one presented at room temperature (Figure 9a), the main diffraction peaks are clearly shifted to slightly lower scattering angle during heating, owing to the thermal expansion of the polymer matrix occurring during heating [56]. At 235 °C, only an amorphous halo was observed, indicating a complete melting of polymer crystals, in agreement with the DSC results (Figure 2a). The subsequent pattern (black curve in Figure 11) was collected at 260 °C, which is the temperature used in DSC experiments to erase the thermal history of pCBT, and, obviously, no diffraction peaks were observed. After melting was completed and the thermal history was properly erased, temperature was decreased in steps and diffractograms were acquired for each step, as reported in blue in Figure 11. During cooling, no crystalline signals appeared down to 220 °C. At 210 °C, low intensity diffraction peaks of pCBT became visible, related to the planes $(0\bar{1}1)$, $(010)$, $(\bar{1}11)$, $(100)$ and $(1\bar{1}1)$, thus evidencing the onset of crystallization. This result is consistent with DSC isothermal crystallization tests, for which 210 °C was the maximum isothermal condition at which crystallization of pCBT was achieved (Figure 3.a). Further decreasing the temperature resulted in the intensification of diffraction patterns and in the slight shift of peaks to higher *2Θ* values, which are related to the completion of pCBT crystallization and to the shrinkage of pCBT unit cell during cooling [56], respectively.

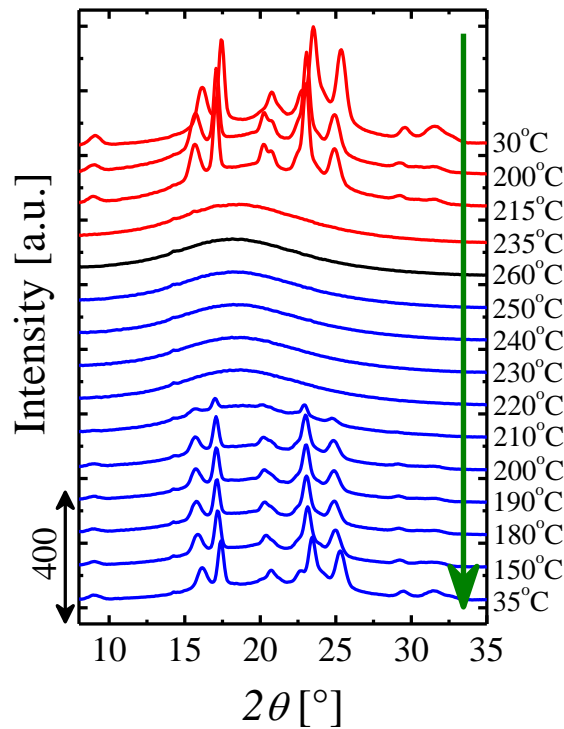

**Figure 11. *In situ* WAXS diffraction patterns collected at different temperatures for pure pCBT. Red and blue curves represent the patterns collected during heating and cooling scans, respectively. The black pattern was collected at 260°C, *i.e.* the temperature at which was erased the thermal history of the material. On the right are reported the temperatures at which was collected each pattern, whereas the arrow indicates the measurement sequence.**

Variable temperature WAXS patterns collected for pCBT + 10% RGO and pCBT + 10% RGO_1700 at different temperatures are reported in Figure 12a and Figure 12b, respectively. Diffraction patterns were collected on heating and cooling scans following the same thermal protocol used for pure pCBT. The red and the blue diffractograms are related to patterns collected during heating and cooling scans, respectively, while the black one represent the scattering pattern recorded at the temperature selected to completely erase the thermal history of pCBT, *i.e.,* 260 °C.

*In situ* measurements on pCBT + 10% RGO (Figure 12a) revealed a similar behavior to that observed for pure pCBT (Figure 11) during heating scans, with the disappearance of peaks related to polymer crystals for temperatures ≥ 235 °C. As expected, the presence of the graphite introduced a new peak located at ~ 26.5° independently on the temperature. During cooling scans, crystallization occur in similar way to that of pCBT, with the simultaneous appearance of the same diffraction peaks related to $(0\bar{1}1)$, $(010)$, $(\bar{1}11)$, $(100)$ and $(1\bar{1}1)$ planes. However, it is worth observing that these peaks appeared at 220 °C, whereas for the pure polymer a higher super cooling (*i.e.,* cooling down to 210 °C) was required for the formation of polymer crystals. This is in agreement with results above reported for isothermal crystallization experiments carried out by DSC (Figure 3.a). However, in WAXS measurements for pCBT + 10% RGO no detectable diffraction signals for the high temperature melting crystalline fraction was observed, even though its presence was detected by non-isothermal DSC, SN and SSA experiments. This is likely due to the extremely low amount of this fraction, ~ 1 J g$^{-1}$ measured by DSC (Table 1), which is probably below the WAXS sensitivity.

Variable temperature WAXS measurements on pCBT + 10% RGO_1700 (Figure 12b and Figure 12c) revealed interesting differences compared to both pCBT and pCBT + 10% RGO. First, the diffraction peak at *2θ* ≈ 26.5°, related to the presence RGO, is clearly more intense respect to that observed in pCBT + 10% RGO (Figure 12a). This is partially explained by the lower defectiveness and higher structural order of RGO_1700, as widely discussed in our previous paper [35], but there may also be an additional effect of higher orientation obtained with annealed RGO. More interestingly, persistence of crystalline organization was found during heating up to 235 °C, thus reflecting the presence of the highly stable crystalline fraction. Upon cooling, at 240 °C traces of diffraction signals appear at *2θ* ≈ 15.9°, 17.1° and 23.1° (indicated by the arrows in Figure 12c), related to the (0-11), (010) and (100) planes, typical of pCBT. When further cooling to 230 °C, all the diffraction peaks related to the pCBT alpha phase were clearly observed, while at 220 °C the peak intensity achieved the maximum value, being the crystallization completed. Comparing these results with those obtained by isothermal crystallization experiments (performed by DSC, Figure 3.a), it appears that the complete crystallization observed by WAXS at 220 °C is in agreement with the maximum isothermal temperature used in DSC tests (219 °C). However, the presence of pCBT diffraction patterns at 235 °C in heating scans and the appearance of the first peaks related to pCBT crystals at 240 °C cannot be regarded as signals related to the main crystallization step of pCBT. Indeed, at these high temperatures only the high stability fraction can survive, in agreement with non-isothermal DSC experiments. Similar results were obtained on pCBT + 50% RGO_1700 prepared by solution mixing (see Supporting information, Figure SI10).

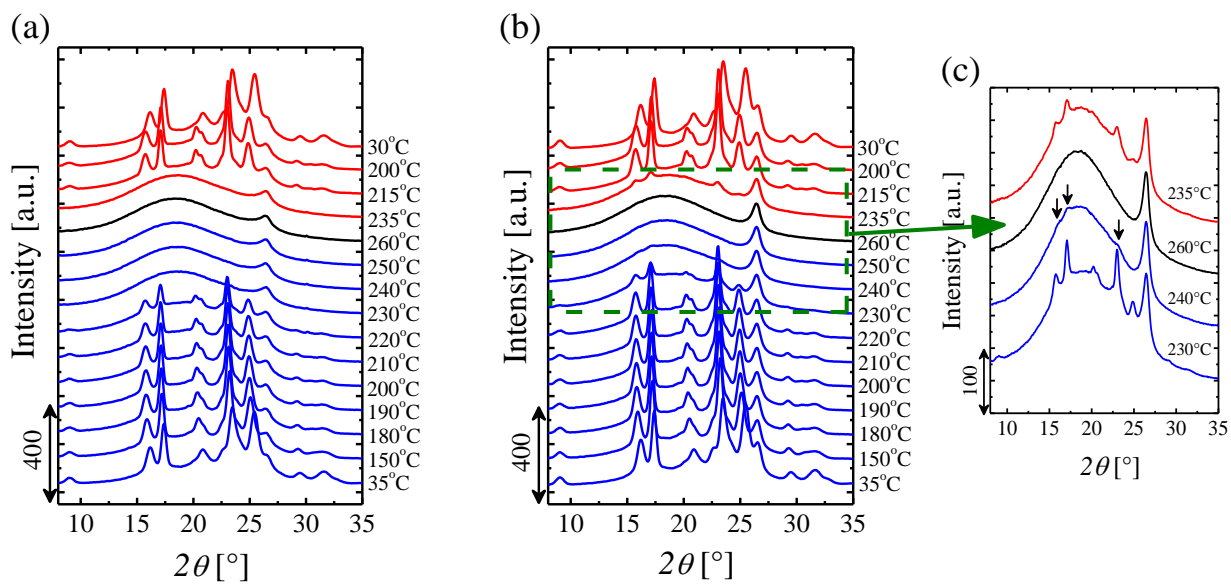

**Figure 12.** *In situ* **WAXS diffraction patterns collected at different temperatures for (a) pCBT + 10% RGO and (b) pCBT + 10% RGO_1700. Selected *in situ* WAXS diffraction patterns for pCBT + 10% RGO_1700 (c). The three arrows (c) indicates the first pCBT crystalline peaks which appear in cooling scans. Red and blue curves represent the patterns collected during heating and cooling scans, respectively. The black pattern was collected at 260°C, *i.e.* the temperature selected to erase the thermal history. On the right are reported the temperatures at which was collected each pattern.**

WAXS results presented here prove that the high temperature melting fraction has the same diffraction pattern observed for the standard pCBT alpha phase. This indicates that the observed higher temperature melting crystal fraction is not constituted by a different crystal phase, hence it must consist of extended chain crystals. Indeed, the measured melting temperature was very close to the equilibrium melting temperature calculated for pure pCBT [15, 20]. This demonstrates, for the first time, the capability of RGO nanoflakes not only to nucleate pCBT but also to induce a very regular arrangement of chains into highly stable crystals, most likely starting their organization from the polymer/nanofiller interface. The importance of such interfacial contact, is further highlighted by the differences between nanocomposites containing RGO and annealed RGO, which demonstrate that structurally ordered and low defective nanoflakes obtained after annealing are much more efficient in promoting the ordered arrangement of polymer chains at the interface. The formation of these small but finite extended chain crystal fraction is also responsible for the supernucleation observed when RGO is added to pCBT as demonstrated by the self-nucleation studies.

## 4. Conclusions

In the present work, pCBT nanocomposites containing 10 wt.% RGO were prepared by ring-opening polymerization of cyclic butylene terephthalate oligomers in the presence of RGO, aiming to study the effect of both conventionally reduced graphene oxide and highly reduced graphene oxide (1700 °C, 1 h, 50 Pa) on the crystallization of linear pCBT.

Organization of RGO nanoflakes in the nanocomposites was assessed by transmission electron microscopy showing homogeneous distribution of the nanoflakes and clear infiltration of the polymer in the expanded structure of nanoparticles, while no significant differences were observed in the morphology of the two nanocomposites prepared with conventionally reduced or highly reduced nanoflakes.

The presence of RGO dramatically affected the crystallization behavior of pCBT. Indeed, non-isothermal DSC showed remarkable shifts of the crystallization peak to higher temperature evidencing a clear nucleating role of the nanoparticles on the crystallization of pCBT. Isothermal DSC experiments showed a strong increase in the crystallization rate of pCBT in nanocomposites, without any alteration of the axialitic superstructural morphology of pCBT crystallization, while the α-crystalline form of pCBT is retained. Self-nucleation experiments revealed that for neat pCBT the three *Domains* of nucleation were clearly recognizable, whereas in nanocomposites *Domain II* was absent. A nucleation efficiency of 164 % and 270 % was calculated for pCBT + 10% RGO and pCBT + 10% RGO_1700, respectively, demonstrating that RGO nanoflakes have a supernucleating effect on pCBT crystallization, *i.e.,* they are better nucleating agents than the polymer self-nuclei. Furthermore, the higher nucleation efficiency for the highly reduced flakes suggests a determinant role of the chemical and physical structure of the graphitic structure on the nucleation of the pCBT crystals. DSC experiments also demonstrated the appearance of a new peak in pCBT nanocomposites having higher enthalpy in the presence of highly reduced graphene oxide (1 J g$^{-1}$ and 4 J g$^{-1}$ for pCBT + 10% RGO and pCBT + 10% RGO_1700, respectively). This peak could be fractionated during SSA experiments, confirming its assignment to a polymer fraction which melted at ~ 250 °C and crystallized at ~ 233 °C. WAXS experiments on pCBT + 10% RGO_1700 showed the persistence of a diffraction pattern at temperatures higher than the standard melting peak of pCBT. This pattern exhibited the same crystalline reflections of pCBT α- form, indicating that the high stability peak is related to a thick stack of pCBT lamellae with a thickness up to 32 nm, calculated according to the Gibbs-Thomson equation. Indeed, 2D-WAXS showed alignment of nanoflakes perpendicularly to the compression direction and orientation of pCBT crystals parallel to the RGO surface, the orientation being stronger in the presence of highly reduced RGO. In particular, the ($1\bar{1}1$) plane of pCBT, which is parallel to the aromatic rings of polymer chains, is the most highly oriented signal in the direction of the RGO flakes, suggesting a self-organization of the pCBT macromolecules driven by π-π interaction between aromatic rings in polymer chains and the sp2 structure of RGO.

It is worth highlighting that this interaction between the nanoflakes and the polymer matrix may be exploited for the engineering of polymer/nanoparticle interfaces in order to improve the related properties for the corresponding nanocomposites, including stress-transfer, heat transfer and gas permeation.

**Authors contributions**

S. Colonna carried out the entire preparation of nanocomposites and most of the characterizations reported in this paper. R.A. Pérez contributed in DSC characterization and A.J. Müller conceived the DSC, SN, TEM and SSA experiments and carried out the interpretation of these results. H. Chen performed X-ray diffraction tests whereas G. Liu and D. Wang carried out XRD interpretation. G. Saracco contributed to the discussion of the results. A. Fina contributed to the interpretation of the results and coordinated the project. Manuscript was mainly written by S. Colonna, A. Fina and A.J. Müller. All authors proof read the manuscript and agreed on the results interpretation.

**Acknowledgements**

The research leading to these results has received funding from the European Union Seventh Framework Programme under grant agreement n°604391 Graphene Flagship. This work has received funding from the European Research Council under the European Union's Horizon 2020 research and innovation programme grant agreement 639495 — INTHERM — ERC-2014-STG. Funding from Graphene@PoliTo initiative of the Politecnico di Torino is also acknowledged. The UPV/EHU team wishes to ackowledge funding by MINECO through grant: MAT2014-53437-C2-2-P. G.L. is grateful to the Youth Innovation Promotion Association of CAS (2015026). The authors gratefully acknowledge the contribution of Dr. Gracia Patricia Leal for her help with TEM measurements.

# SUPPORTING INFORMATION

*Standard DSC on pCBT + 50% RGO_1700 prepared by solution mixing*

To investigate the formation of the high stability crystals, increase of the amount of this crystalline population was attempted via solution mixing, which allowed preparing a pCBT composite containing 50% wt. of RGO_1700, which is clearly not feasible via melt compounding.

DSC heating and cooling scans for pCBT + 50% RGO_1700 after solvent evaporation are reported in Figure SI 1. In the first heating, the main melting peak of pCBT was characterized by a shoulder in the low temperature side indicating imperfect wide distribution of crystal size and/or defectiveness, obtained during solvent evaporation. Furthermore, no clear signs of the highly stable crystals were observed. However, in cooling scans, and in the subsequent heating scans, the high temperature crystalline population becomes clearly visible, at slightly lower temperatures (crystallization at ~ 224 °C and melting at ~ 241 °C) compared to the melt-processed nanocomposite (containing a lower amount of nanoflakes). The crystallinity was ~ 60 %, whereas the enthalpy related to the highly stable crystal fraction was about 12 % of the main peak value, thus confirming an increase compared to pCBT + 10% RGO_1700, in which the high stability fraction corresponds to about 7 %. Clearly, the increase of this new crystalline population is non-linear with increasing the nanoflakes content, which may be due to different possible explanations. On the one hand, it is likely that the limit for the crystallization into a highly stable fraction is related to the dispersion of the nanoparticles, thus affecting the extent of interfacial surface area. With a nanoparticle loading of 50 wt.%, aggregation of nanoflakes certainly occurs. This leads to an effective interfacial area, with the polymer, which is expected to be far below the maximum theoretical value, limiting the influence of the nanoparticles on the crystallization of pCBT in proximity of the same nanoflakes. On the other hand, there may also be an effect of shearing: a relatively high shear rate is applied during extrusion and polymerization of CBT into pCBT, which may lead to an intimate contact between the polymer and the nanoflakes, as a consequence of chains orientation, while no significant shear is applied during solution mixing.

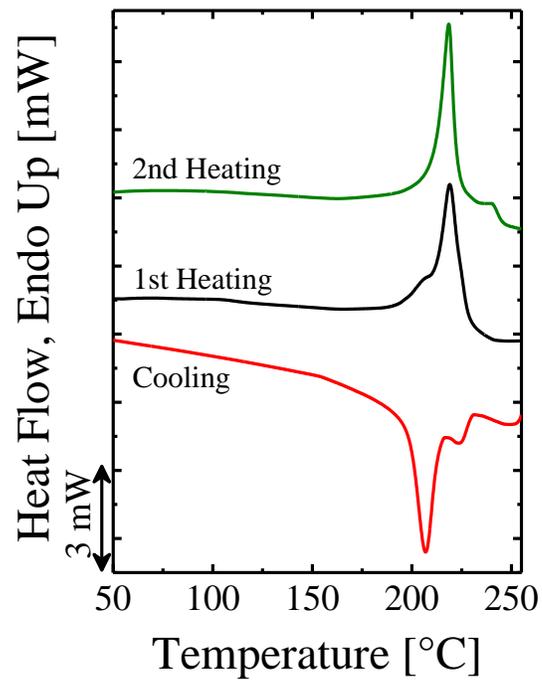

**Figure SI 1. Standard DSC cooling, 1st and 2nd heating scans for pCBT + 50% RGO_1700 obtained by solvent mixing**

*Thermal stability*

Non-isothermal DSC cooling scans were employed to study how the temperature selected to erase the thermal history of the sample ($T_{max}$), affects $T_c$. For this reason, samples were heated from 25 °C up to $T_{max}$ (at 20 °C min$^{-1}$), held at $T_{max}$ for 3 minutes, then cooled down to 25 °C (at 20 °C min$^{-1}$) and held at this temperature for 1 minute. This procedure was repeated for 26 times. Tests were carried out only on pure pCBT and new samples were used for each test. Experimental results for $T_c$ *vs.* N, obtained after tests at the different $T_{max}$, are summarized in Figure SI 2.

$T_c$ *vs.* N plots, reported in **Errore. L'origine riferimento non è stata trovata.**, show that the crystallization peak temperature generally increases as the number of cycle increase, independently on $T_{max}$, thus indicating a degradation of the polymer matrix, with a reduction of the average pCBT chain lengths. Indeed, for entangled systems a decrease in the molecular weight can result in an initial increase in the crystallization temperature, owed to a higher mobility of polymer chains. It is noteworthy that the evolution of $T_c$ against cycles is strictly connected to the set $T_{max}$. For $T_{max}$ = 250 °C and $T_{max}$ = 260 °C, the measured $T_c$ started at about 189 ± 0.5°C and monotonically increase with cycles, with a higher slope at the higher temperature. When heating at $T_{max}$ = 280°C, crystallization temperature on first cycle was measured at about 191 °C and raply increased with N, indicating a fast and extensive thermal degradation upon cycling. The limited thermal degradation observed when $T_{max}$ = 250 °C made the use of this temperature interesting as upper limit for SN and SSA experiments. However, the presence of the high temperature melting phase (~ 250°C) in pCBT nanocomposites would result in the presence of crystal fragments which could act as nucleating agent, affecting the experiments. For this reason, for SN and SSA test was selected $T_{max}$ = 260 °C as a good compromise between thermal degradation and erasure of the thermal history.

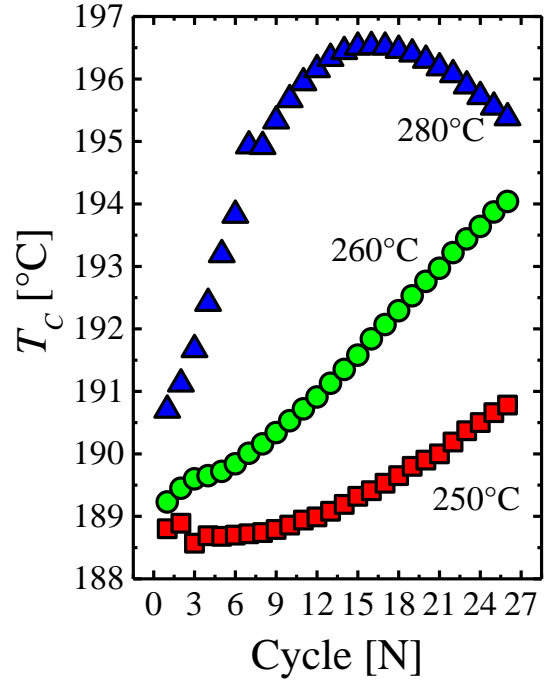

**Figure SI 2.** Tc measured in standard cooling scans after heating up to Tmax (selected to erase the thermal history and reported in the graph) for N times

*Isothermal crystallization*

**Table SI 1. Parameters obtained by fitting with Avrami theory the data obtained from isothermal crystallization tests.**

| sample | $T_c$ [°C] | n | $K \times 10^3$ [min$^{-n}$] | $K^{1/n}$ [min$^{-1}$] | $\tau_{1/2t}$ [min] | $\tau_{1/2e}$ [min] | $R^2$ |
|---|---|---|---|---|---|---|---|
| pCBT | 205 | 2.1 | 269.0 | 0.54 | 1.56 | 2.11 | 0.9995 |
|  | 206 | 2.0 | 174.0 | 0.42 | 2.01 | 2.22 | 0.9995 |
|  | 207 | 2.0 | 60.7 | 0.25 | 3.43 | 3.88 | 0.9998 |
|  | 208 | 2.0 | 28.2 | 0.17 | 5.17 | 5.53 | 1 |
|  | 209 | 2.0 | 11.2 | 0.11 | 8.03 | 8.52 | 0.9999 |
|  | 210 | 1.9 | 7.2 | 0.07 | 10.50 | 10.69 | 0.9999 |
| pCBT + 10% RGO | 210.5 | 1.8 | 91.5 | 0.26 | 3.18 | 3.57 | 0.9998 |
|  | 211 | 1.7 | 78.0 | 0.22 | 3.61 | 4.19 | 0.9997 |
|  | 211.5 | 1.7 | 52.6 | 0.18 | 4.51 | 5.43 | 0.9994 |
|  | 212 | 1.7 | 39.8 | 0.15 | 5.44 | 6.32 | 0.9998 |
|  | 212.5 | 1.7 | 34.4 | 0.14 | 5.95 | 6.85 | 0.9993 |
|  | 213 | 1.7 | 25.0 | 0.11 | 7.01 | 7.59 | 0.9999 |
| pCBT + 10% RGO_1700 | 218 | 1.7 | 23.5 | 0.11 | 7.70 | 8.83 | 0.9994 |
|  | 218.5 | 1.6 | 18.4 | 0.08 | 9.18 | 9.97 | 0.9990 |
|  | 219 | 1.5 | 17.2 | 0.07 | 11.52 | 12.54 | 0.9993 |

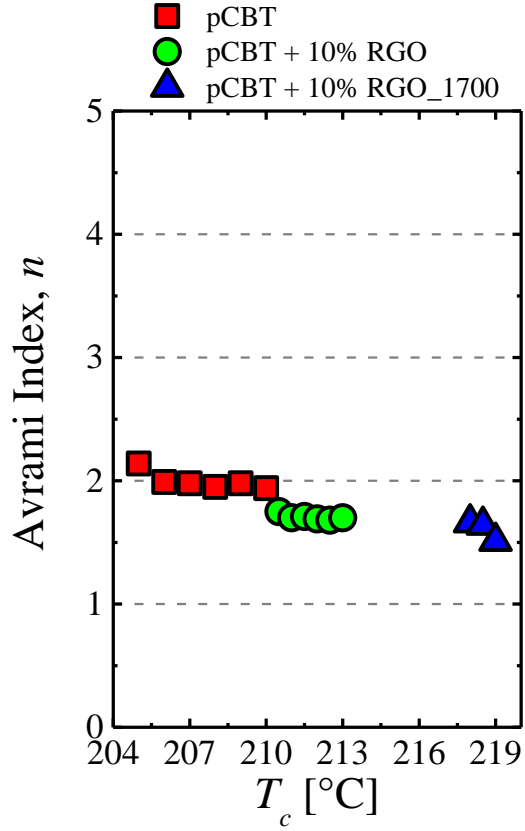

**Figure SI 3. Avrami index, *n*, as a function of isothermal crystallization temperature for pCBT and pCBT/rGO nanocomposites**

**Table SI 2. Parameters obtained from fitting the Lauritzen and Hoffman to the data of Figure 2a**

| sample | $K_g^\tau$ [K²] | $\sigma$ [erg cm⁻²] | $\sigma_e$ [erg cm⁻²] | $q \times 10^{13}$ [erg] | $R^2$ |
|---|---|---|---|---|---|
| pCBT | 517000 | 10.60 | 208.9 | 12.1 | 0.9776 |
| pCBT + 10% RGO | 374000 | 10.60 | 151.5 | 8.8 | 0.9615 |
| pCBT + 10% RGO_1700 | 306000 | 10.60 | 123.8 | 7.2 | 0.9750 |

*Self-nucleation*

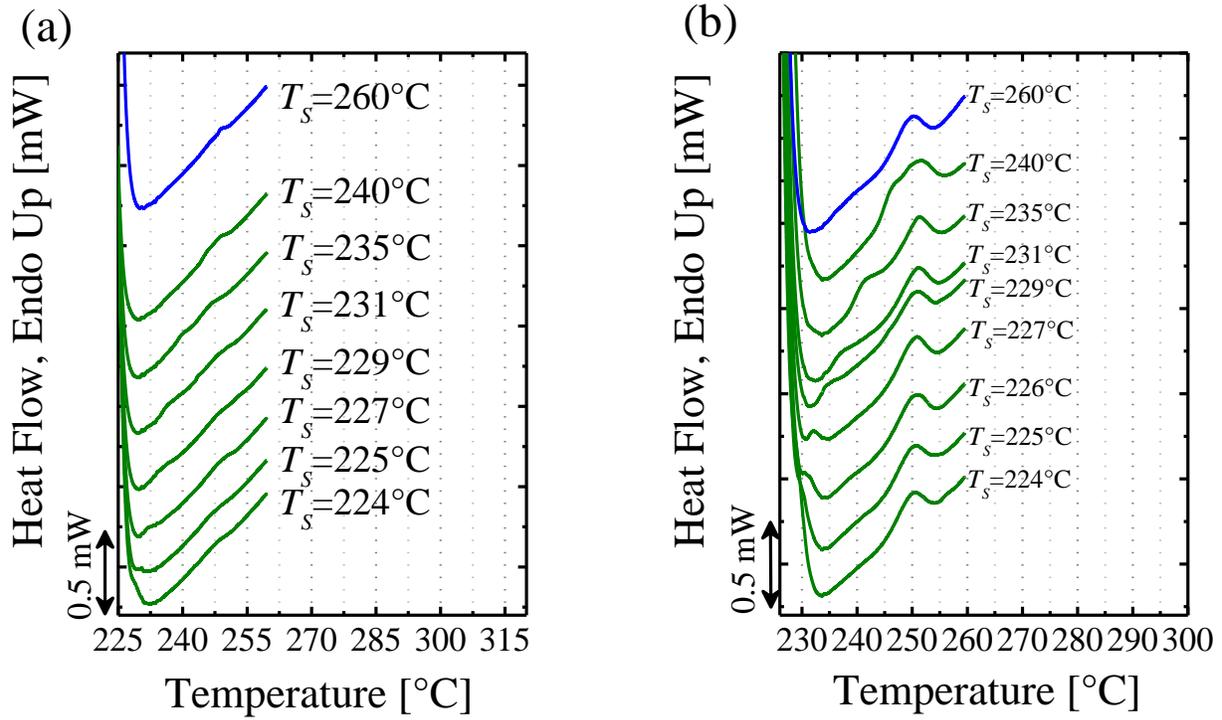

**Figure SI 4.** Magnification at high temperature on DSC heating scans performed after cooling from the indicated $T_s$ for (a) pCBT + 10% RGO and (b) pCBT + 10% RGO_1700

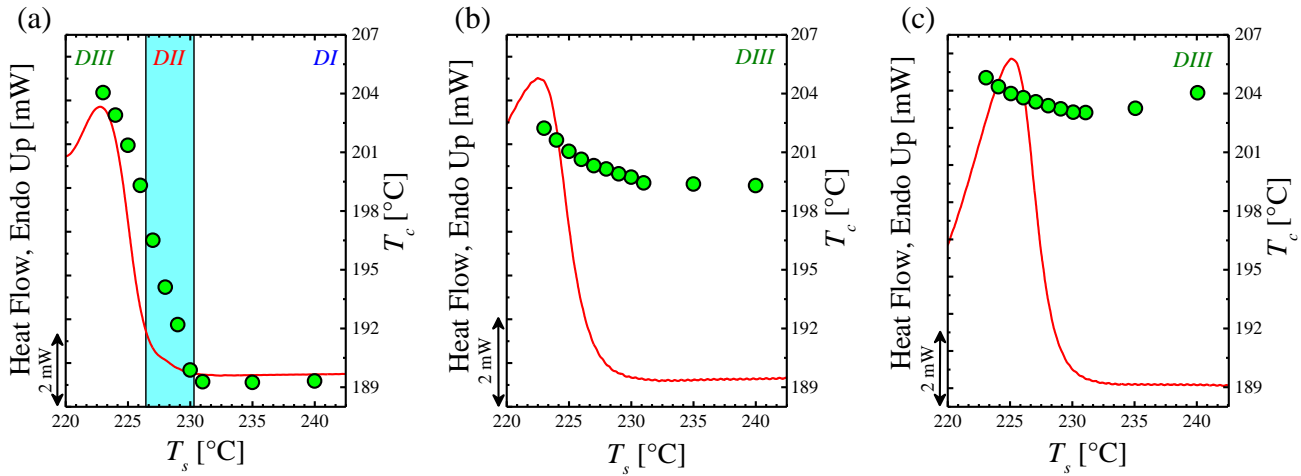

**Figure SI 5.** Standard DSC heating scans (red line) plotted along with crystallization peak temperatures (green circles) *vs.* $T_s$ for (a) pCBT, (b) pCBT + 10% RGO and (c) pCBT + 10% RGO_1700. The vertical lines indicate the *Domain* borders. The temperature range at which materials experienced *Domain II* is highlighted.

*Successive self-nucleation and annealing*

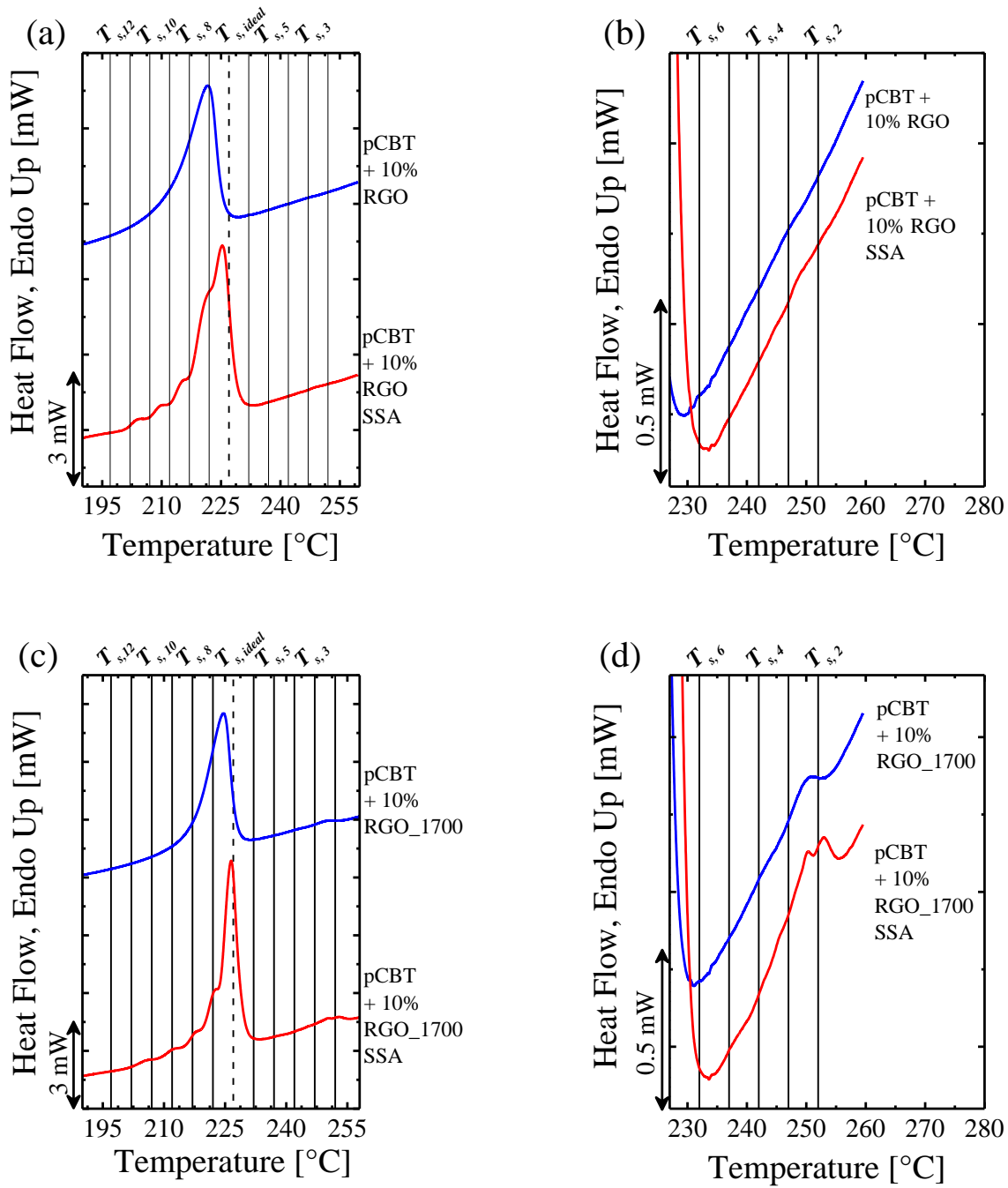

**Figure SI 6.** DSC heating scans for (a) pCBT + 10% RGO and (c) pCBT + 10% RGO_1700, before (blue curves) and after (red curves) SSA thermal fractionation. The effect of thermal fraction on the high temperature melting phase is reported in (b) for pCBT + 10% RGO and (d) for pCBT + 10% RGO_1700. The solid vertical lines represented the values of $T_s$ temperature employed for thermal fractionation while the dashed vertical line indicates the $T_{s,ideal}$ for pCBT.

*Wide Angle X-Ray Scattering*

**Figure SI 7.** Intensity distribution of selected reflections *vs.* azimuthal angle ($\phi$) collected perpendicularly to the compression direction for pCBT + 10% RGO_1700

**Figure SI 8.** 2D WAXS patterns measured via transmission geometry on pCBT (a) parallel and (b) perpendicular to the compression direction.

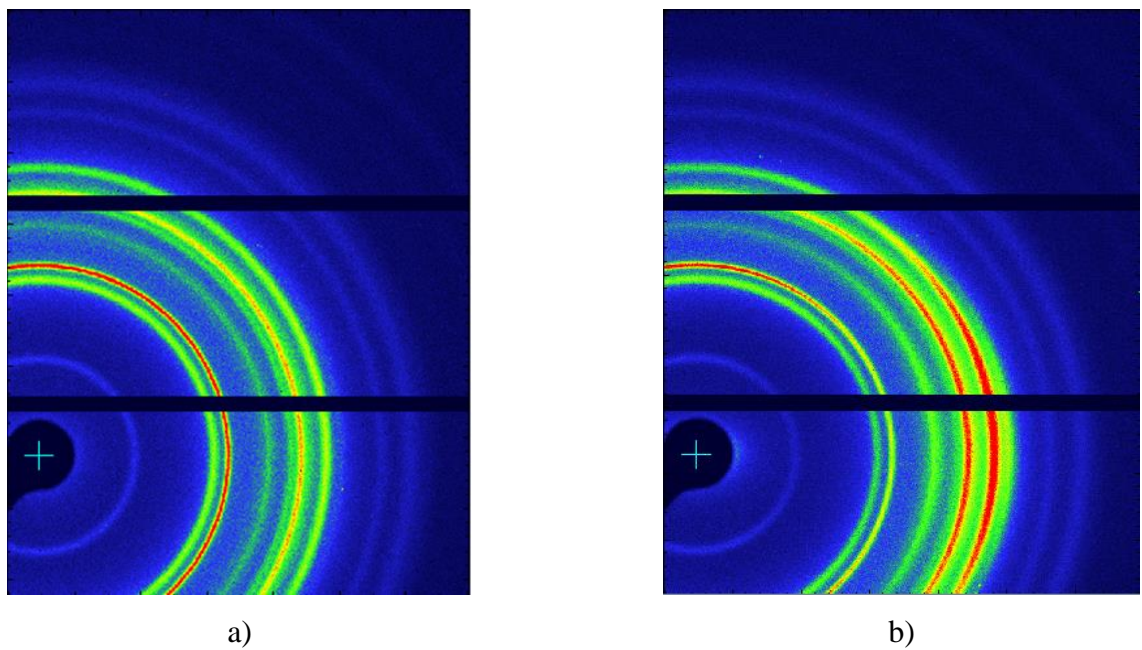

a)                                                                                           b)

**Figure SI 9. 2D WAXS patterns measured via transmission geometry on pCBT + 10% RGO (a) parallel and (b) perpendicular to the compression direction.**

*Temperature assisted WAXS*

Trying to further characterize the high melting phase, WAXS were performed on pCBT + 50% RGO_1700 prepared by solution mixing and hot compressed into a film. Results (Figure SI 10) are well consistent and similar with those described for pCBT + 10% RGO_1700 even if the first crystalline peak related to pCBT crystals, ~ 17.0 ° (010), are observed at 230 °C despite the general lower intensity of pCBT diffraction pattern respect to that of pCBT + 10% RGO_1700.

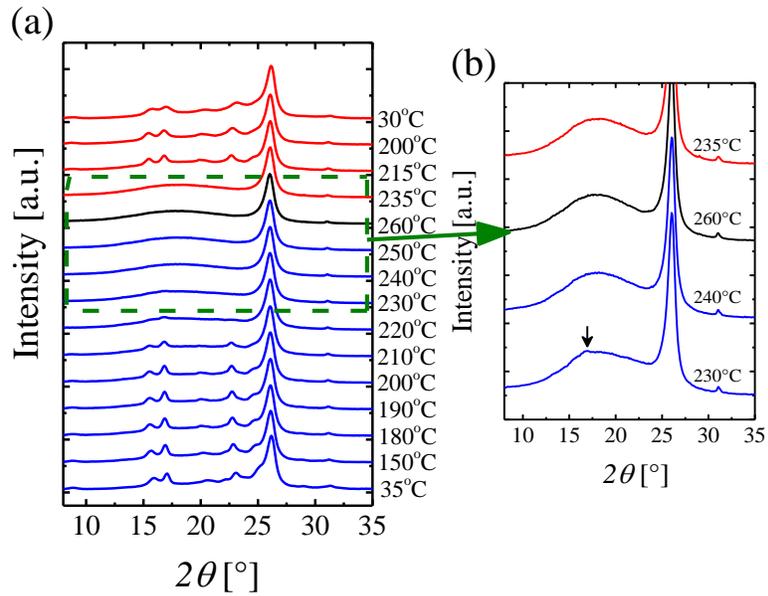

**Figure SI 10.** In situ WAXS diffraction patterns collected at different temperatures for pCBT + 50% RGO_1700 (a) and selected diffraction patterns (b). The arrow in (b) indicates the first pCBT crystalline peak which appears during cooling scans. Red and blue curves represent the patterns collected during heating and cooling scans, respectively. The black pattern was collected at 260°C, *i.e.* the temperature selected to erase the thermal history. On the right are reported the temperatures at which was collected each pattern.